\documentclass[]{elsarticle01}

\usepackage{amsmath, amssymb}

\newtheorem{proposition}{Proposition}
\newtheorem{theorem}{Theorem}
\newtheorem{lemma}{Lemma}

\newtheorem{definition}{Definition}

\def\beginProof{\par{\it Proof}. \ignorespaces}
\newcommand{\closeProof}{\hfill $\square$}

\newcommand{\beginE}{\begin{equation}}
\newcommand{\closeE}{\end{equation}}

\newcommand{\ds}{\displaystyle}
\newcommand{\realR}{{\mathbb{R}}}

\newcommand{\inner}[2]{\big\langle #1, \, #2 \big\rangle}
\renewcommand{\outer}[2]{\big(  #1, \, #2 \big) }

\newcommand{\myand}{\qquad \mbox{and} \quad}
\newcommand{\myor}{\qquad \mbox{or} \quad}

\newcommand{\E}{\mathrm{E}}
\newcommand{\VAR}{\mathrm{VAR}}
\newcommand{\COV}{\mathrm{COV}}
\newcommand{\ISdlr}{\mathrm{IS}_{\$}}
\newcommand{\ISbps}{ \mathrm{IS}_{\mathrm{bps}} }
\newcommand{\bps}{\mathrm{bps}}
\newcommand{\sign}{\mathrm{sign}}
\newcommand{\side}{\mathrm{side}}
\newcommand{\maxPoV}{\mathrm{maxPoV}}
\newcommand{\opI}{\mathbf{I}}
\newcommand{\opK}{ \mathbf{K}_\lambda  }
\newcommand{\opD}{\mathbf{D}}

\newcommand{\sprd}{\mbox{IC-sprd}}
\newcommand{\inst}{\mbox{IC-inst}}
\newcommand{\perm}{\mbox{IC-perm}}
\newcommand{\tran}{\mbox{IC-tran}}

\newcommand{\NBO}{\mathrm{NBO}}

\newcommand{\eps}{\varepsilon}

\begin{document}

%% ---------------  START:  Title Page -------------->>>
\begin{frontmatter}

\title{  A Pre-Trade Algorithmic Trading Model under 	% \\
		 Given Volume Measures and Generic Price Dynamics ({\small GVM-GPD}) }
	
\author{ Jackie (Jianhong) Shen \fnref{fn_shen2} }
\ead{jhshen@illinois.edu}
\ead[url]{http://publish.illinois.edu/jackieshen}
\address{ 
		Industrial and Enterprise Systems Engineering (ISE)    \\
        and Master of Science in Financial Engineering (MSFE) \\
        216D Transportation Building, University of Illinois, Urbana, IL 61801 \\ 
        \vskip 10pt
        {\sf Dedicated to Gil Strang on the Occasion of His 80th Birthday.}
}
						
\fntext[fn_shen2]{Jackie Shen had been a Vice President for the Equities Algorithmic Trading divisions at both J.P. Morgan and Barclays, New York, before joining the University of Illinois in 2013. Readers are encouraged to first read the disclaimer statements near the end.}

\begin{abstract}
We make several improvements to the  mean-variance framework for optimal pre-trade algorithmic execution, by working with volume measures and generic price dynamics. Volume measures are the continuum analogies for discrete {\em volume profiles} commonly implemented in the execution industry. Execution then becomes an {\em absolutely continuous} measure over such a measure space, and its Radon-Nikodym derivative is commonly known as the  {\em Participation of Volume} (PoV) function. The four impact cost components are all consistently built upon the PoV function.  Some novel efforts are made for these linear impact models by having market signals more properly expressed.  For the opportunistic cost, we are able to go beyond the conventional Brownian-type motions. By working directly with the auto-covariances of the price dynamics, we remove the Markovian restriction associated with Brownians and thus allow potential memory effects in the price dynamics.  In combination, the final execution model becomes a constrained quadratic programming problem in infinite-dimensional  Hilbert spaces. Important linear constraints such as participation capping are all permissible. Uniqueness and existence of optimal solutions are established via the theory of positive compact operators in Hilbert spaces. Several typical numerical examples explain both the behavior and versatility of the model.

\end{abstract}

\begin{keyword}
Volume \sep Price \sep Impact \sep Risk \sep Compact positive operator \sep Hilbert space \sep Existence \sep Uniqueness \sep Quadratic Programming
\end{keyword}

\end{frontmatter}
%% <<<---------------  END:  Title Page ----------------

%% Main Sections
\section{Introduction}
Algorithmic trading helps institutional investors liquidate or acquire big positions without incurring much adverse impact cost or opportunistic risk. To achieve this objective, a practical model must keep screening any real-time market characteristics and adapt the execution strategy accordingly. This necessarily means that a realistic trading model has to be {\em dynamic}, as in the seminal work of Bertsimas and Lo~\cite{algo_bertsimas_Lo98}, and many other important ones (e.g., Huberman and Stanzl~\cite{algo_huberman_stanzl01}, Almgren~\cite{algo_almgren12}, Bouchard, Dang, and Lehalle~\cite{algo_bouchard_dang_lehalle2011}, Azencott et al.~\cite{algo_azencott_lehalle_etal2013}, just to name a few).

On the other hand, both internal or external clients commonly rely on computationally less expensive {\em static} models to get pre-trade estimations on potential costs and risks for their positions. They may also rely on such pre-trade estimations to generate alerts if the actual execution is deviating too far, often expressed via confidence intervals, and needs immediate human intervention. Furthermore, due to the mounting computational challenges associated with fully dynamic algorithms, execution houses (e.g., broker-dealers, agency services, internal execution teams in hedge funds, etc.) also typically utilize {\em static algorithms} as the core for building heuristic but much faster dynamic models.

It is for these reasons that the study of utility-function based {\em static } algorithms is still very valuable in the execution industry. The current work focuses on some major improvements of the mean-variance framework, following largely the classical works of Almgren and Chriss~\cite{algo_almgren_chriss00}, Huberman and Stanzl~\cite{algo_huberman_stanzl01}, and other closely related works (e.g., Kissell and Malamut~\cite{algo_kissell_malamut06}, and Obizhaeva and Wang~\cite{algo_obizhaeva_wang13}).

The core of the current work is a single continuum model independent of any interval based discrete grids. This does not mean that any in-house implementation has to avoid interval-based setting. Instead, a single governing continuum model has numerous advantages, including for example, (i) ensuring consistency among different interval choices, (ii) staying invariant when technology advancements (e.g., on data servers, optimizer servers, or Direct Market Access (DMA)) allow executing in higher and higher frequencies, and (iii) welcoming more computational methodologies, including for example, basis function based methods that are free of interval grids. Imaging Science, for instance, has witnessed a blossoming decade very much thanks to the similar advantages as one goes from pixel or graph based discrete models to continuum models, as the latter ones are independent of camera resolutions and also befriend a wealth of mathematical tools such as PDE, variational calculus, and wavelets (e.g., Chan and Shen~\cite{bImage_ChanShen}). 

The two major characteristics of the proposed model are, as the title has suggested, (i) treating market volumes as Borel measures over the execution horizon, and (ii) permitting generic price dynamics (including generic Bid-Ask spread dynamics), other than being restricted to the linear or geometric Brownians prevailing in the literature. 

Generic price dynamics extend beyond the Markovian nature of conventional Brownian-type price movements, and allow short-term or long-term memories. This will be explained in more details in Section~\ref{S:price_dynamics}. Two more specific examples involving mean reversion and stochastic volatilities are also presented in Section~\ref{S:computation} to illustrate the flexibility of the model. 

In the typical reductionism approach to static modeling of algorithmic trading~\cite{algo_almgren_chriss00,algo_obizhaeva_wang13}, the markets have usually been approximated by the volume profile  (instead of more complex signals associated with the dynamics of {\em limit order books} (LOB)). Volume measures generalize  discrete {\em volume profiles} to the continuum setting. We attempt to demonstrate that measure theory and functional analysis serve a natural foundation for the continuum modeling. Section~\ref{S:volume_measures} lays out the general setting for volume measures. 

The proposed execution model is built upon four impact cost components, which are all {\em linear}. This is the third characteristics of the current work. Linear cost models allow faster calibration via linear regression techniques. They also lead to {\em Quadratic Programming} (QP) formulation for the final execution model, which can be readily solved via robust commercial QP solvers (e.g., the IBM CPLEX). With volume measures, all the four impact cost components are consistently build upon the {\em Participation of Volume} (PoV) rate function, which is the Radon-Nikodym derivative of the execution measure against the market volume measure.  For both the transient and permanent costs, we have introduced volume distance for the transient impact and cumulative volume denomination for the permanent impact. These efforts are novel to the best knowledge of the author, and make the cost models more realistic. Throughout the model building process, we have also particularly emphasized:
\begin{enumerate}[(a)]
\item the influence of in-house {\em Child Order Placement} strategies on modeling parameters, and
\item the role of quantitative market makers on shaping the exact forms of impact models. 
\end{enumerate}
All these subjects will be further elaborated in Section~\ref{S:costcomponents}. 

The rest of the paper is devoted to the analysis and computation of the established model. In Section~\ref{S:model_analysis}, we apply the theory of positive compact operators in Hilbert Spaces to establish the existence and uniqueness of solutions to the resulted QP problem with a quadratic objective and linear constraints. A computational scheme is then proposed in Section~\ref{S:computation}, and several numerical examples are also presented to reveal the effects of various factors, including degrees of risk aversion, shapes of volume measures, dominance of individual cost components, and non-Brownian price dynamics. A feature that is also somewhat novel in the proposed model is the explicit permission of volume capping, which is popularly demanded from both internal and external clients.

\section{Generic Price Dynamics} \label{S:price_dynamics}

Regardless of execution styles, trading is always exposed to the overall market movements. This is especially true for large institutional orders for which immediate filling via open exchanges is infeasible. In the current work we will not touch the subject of {\em dark pools} where it is {\em not} impossible to have a large order filled via internal crossing networks run by broker-dealers, agency services, or even exchanges.

This part of the trading cost is often termed the {\em opportunistic cost or risk}, and is directly caused by the innate price fluctuations of the target securities. Trading models have to incorporate suitable price dynamics in order to quantify the associated opportunistic cost.  

Let $p_0$ denote the arrival price at time $t=0$ on the normalized trading horizon $I=[0, 1]$. Practically, this could be the mid-quote of the {\em National Best Bid and Offer} (NBBO). The mid-quote of the NBBO has been traditionally considered as the {\em true} price of the security~\cite{algo_obizhaeva_wang13}, an assumption we shall follow as well. Let $p_t$ or $p(t)$ denote the mid-quote at time $t \in I$. Define the price fluctuation to be $\delta_t = p_t - p_0$, or its homogenized version to be: $\delta_t^\circ = \delta_t / p_0$. It has been commonly assumed in the literature that without impact costs,  $\delta_t$ is either linearly or geometrically Brownian~\cite{algo_almgren_chriss00,algo_obizhaeva_wang13}, i.e.,
\[
    d\delta_t = \hat \mu dt + \hat \sigma dW_t, \myor d \delta_t /p_t = \mu dt + \sigma dW_t.
\]
The model in the current work does not specifically address the overnight risks, and is primarily designed for intraday executions. For the intraday horizon, the geometric Brownian can be well approximated by the linear substitute:
\[
    d \delta_t /p_0 = d \delta^\circ_t = \mu dt + \sigma dW_t.
\]
This works well for most {\em non-penny} common stocks whose prices are sufficiently positive (e.g., above \$5.00),  and intraday price changes are no more than a few percentage points so that the dynamic denominator $p_t$ can be well approximated by the static reference price $p_0$.

The current work, however, makes no assumptions on the specific format of the price dynamics. Instead, we only rely on the auto-covariance function: for any $t, s \in I=[0, 1]$,
\[
    K_\delta^\circ(t, s) = \E\left[  \delta^\circ_t \cdot \delta_s^\circ \right], 	\myor K_\delta(t,s)  = \E \left[ \delta_t \cdot \delta_s \right]
    \; \; \left(=p_0^2 K_\delta^\circ(t, s)\right) .
\]
For the intraday horizon, empirical evidences show that the drifting component $\mu$ in the Brownian framework can be assumed zero. We shall also assume that $\delta_t$ (or $\delta_t^\circ$) is a zero-mean process. 

For the {\em kernel} function $K_\delta(t, s)$, the following natural assumptions are to be made:
\begin{enumerate}[(i)]
\item (Symmetry) $K_\delta(t, s)= K_\delta(s, t)$, for any $t, s \in I=[0, 1]$.
\item (Positivity) for any {\em finite} subset $\Lambda \subset I$, and any real function on the set:
    \[
        a: \Lambda \to (-\infty, +\infty), \qquad t \to a_t,
    \]
     one always has:
     \[
            \sum_{t, s \in \Lambda} K_\delta(t, s) a_t a_s \ge 0.
     \]
\end{enumerate}

Conventional Brownian models always satisfy these conditions with the kernel: 
\[
    K_\delta(t, s) = p_0^2 \cdot \sigma^2 \cdot \min( t, s) =  p_0^2 \cdot \sigma^2 \cdot  (t \wedge s).
\]
(For convenience, we denote $\min(t, s)$ by $t \wedge s$, and $\max(t, s)$ by $t \vee s$.) Furthermore, by working with the auto-covariance kernel alone, the underlying price dynamics do not need to be Markovian or memoryless, which provides much more flexibility in applications. In Section~\ref{S:computation}, we have implemented two non-Brownian examples whose price dynamics are governed by mean reversion or stochastic volatility.

\section{Volume Measures and PoV} \label{S:volume_measures}

Besides the extension beyond the conventional Brownian-type price dynamics, another major methodological innovation in the current work is we assume that the market volume distribution $dV$ over the execution horizon $t \in I=[0, 1]$ is a finite and positive Borel measure.

Let $dt$ denote the traditional  Lebesgue measure over the horizon $I$. If the volume measure is   Lebesgue {\em absolutely continuous}, the market execution {\em speed}  $v(t)$ is then well defined to be the Radon-Nikodym derivative:
\[      v(t) = \frac{ dV}{dt}, \myand v(t) \in L^1(I) = L^1(I, dt).     \]
Generally, embedded within a  Borel volume measure could also be  {\em atomic} measures and  {\em continuous} singular measures. In particular, the latter means that market volumes could grow {\em  stealthily} on a set of times whose total Lebesgue measure is zero. This could be particularly useful when studying the joint effects of {\em lit and dark} pools, though we will not expand this topic in the present work.  

Suppose a client intends to execute $X_1$ shares of some security, over a normalized time interval of $I=[0, 1]$.  Set $X_1$ positive if it is a buy to open a long position or to close an existing short position, and negative if it is a sell to liquidate an existing long position or to open a new short position. Since the current model assumes symmetry between buys and sells, from now on we shall only work with a {\em buy} order as the default setting, unless stated otherwise.

The {\em pre-trade} execution is then considered to be a Borel measure $dX$ over the horizon $I$. And the following basic assumptions are made throughout the work.
\begin{enumerate}[(1)]
  \item (Monotone) $dX$ is a positive measure for buys (and negative for sells). This is a basic request from clients who are generally against opportunistic selling for a buy order, and vice versa. This implies that the cumulative shares $X_t$ or $X(t)$ always change monotonically from $0$ to $X_1$.
  \item (Completion) $ \int_I dX = X_1$, i.e., the targeted order is entirely filled during the horizon. 
  \item (Absolute Continuity) We assume that $dX$ is {\em absolutely continuous} with respect to the market volume measure $dV$, so that the {\em Participation of Volume} (PoV) rate function $h(t)$ is the Radon-Nikodym derivative:
        \[
            h(t) = \frac{ dX } { dV }, \myand h \in L^1(I, dV).
        \]
        The monotone assumption now simply states that $h$ never changes sign (almost every (a.e.) with respect to $dV$). Similarly, the completion condition becomes
        \[
             \int_I h(t) dV(t) = X_1.
        \]
\end{enumerate}

Recall that in measure theory~\cite{evagar,fol}, the absolute continuity condition is equivalent to requiring that on any time subset $\Gamma \subseteq I$, $\ds V(\Gamma)=\int_\Gamma dV = 0$ would imply $\ds X(\Gamma)= \int_\Gamma dX = 0$. We call this the {\em fundamental principle of trading}. In the reductionism approach the market volume alone represents the entire market. Traders therefore would naturally avoid trading whenever there are no market activities.

In what follows, we shall consistently build all the cost and risk components upon the PoV rate function $h(t)$ or $h_t$, which then becomes the decision variable to optimize with.

\section{Components of Linear Impact Models}
\label{S:costcomponents}

The impact model in the current work includes four pieces of components: spread cost, instantaneous cost, transient cost, and permanent cost.
The first two components contribute to realized impact cost, but leave no trailing imprints on market prices, while the latter two do. All components are not unfamiliar in the literature (e.g.,~\cite{algo_bertsimas_Lo98,algo_almgren_chriss00}), but some  innovation efforts have been made:
\begin{enumerate}[(i)]
\item For the spread cost, stochastic spreads are allowed, and the component coefficient is explicitly linked to the {\em Child Order Placement} of each execution house.

\item  For the transient cost, we rely on a quantitative market-maker mechanism to build a volume-distance based exponential transition model.  This effort is very much inspired by the temporal exponential resilience of Obizhaeva and Wang~\cite{algo_obizhaeva_wang13}, whose driving market force has however not been elucidated. 

\item For the permanent cost, we employ the look-back PoV as the dependent variable, instead of the cumulative trading volume commonly used in the literature.  This allows to differentiate the permanent costs arising from different market volume environments, e.g., those due to trading 1000 shares when the market volume is 5,000 shares versus when the market volume is 20,000 shares.
\end{enumerate}

Furthermore, in the current work, we shall also stick to {\em linear} models for all the components, which result in a quadratic programming problem in Hilbert Spaces for the ultimate trading model. Another advantage of linearity is that model calibration could directly rely  upon linear regression (see Section~\ref{secsub:calibration}), instead of nonlinear solvers which are slower or with no guarantee on uniqueness.  

\subsection*{Spread Cost: $\sprd$}

Let $\theta_t$ denote the NBBO spread at time $t$ (in dollar amount), and $\theta_t^\circ = \theta_t/p_0$ the homogenized spread with respect to the reference arrival price. That is,
\[
    \theta_t = \mathrm{NBO}_t - \mathrm{NBB}_t,
\]
the gap between the national best offer (NBO) and best bid (NBB).
For instance, for the S\&P-500 (large-cap) universe, the median homogenized spread is about 5.5 basis points, i.e., $\theta_t^\circ$ roughly fluctuates around $5.5/10,000= 0.00055$. For the S\&P-400 (mid-cap) universe, it roughly doubles.

Since the mid-quote of the NBBO is considered as the {\em true} price, as one crosses the spread to consume market liquidity, one has to pay at least the half spread. More generally, we denote the signed spread cost (per share) by:
 \[
    \sprd_t = \alpha_0 \cdot \sign(h_t) \cdot \theta_t.
 \]
Here the sign of the PoV indicates to pay higher for buying and receive lower for selling. Typically the coefficient $\alpha_0$ is somewhere between 0.0 and 0.5. Regression results from real trades in some of the execution houses where the author has worked have confirmed this general behavior.

There is practical reason why in reality $\alpha_0$ is less than $0.5$ and clients on average pay less than the half spread. Each trading model is typically implemented on a discrete time grid with interval length equal to, say, 1 minute or 5 minutes (often pegged to the liquidity level of the target security). In the modern environment of high-frequency technology, each interval period provides sufficient time to make smart {\em Child Order Placement} (COP). This typically involves a mixture of {\em limit} and {\em market} orders. In reality, COP could also access internal or external non-displayed crossing networks, or the {\em dark pools}.  Successful limit orders eliminate the need to cross the half spread to take liquidity as market orders do, and actually gain the half spread. Most dark orders, on the other hand, are usually pegged to the mid price of the NBBO and are thus traded at the {\em true} price without spread gain or loss. In combination, realized spread cost is typically less than the half spread. 

Therefore, the coefficient $\alpha_0$ partially reflects the underlying COP strategy. Different houses may observe different $\alpha_0$ due to different COP strategies that are  tailored to their accessible trading venues  and  proprietary anti-gaming strategies. 

To proceed, we further assume  that $\theta_t$ has a mean process $\bar \theta_t$, and auto-covariance kernel $K_\theta$:
\[
    K_\theta(s, t) = \COV( \theta_s, \theta_t ) = \E\left[ (\theta_s- \bar \theta_s)(\theta_t - \bar \theta_t) \right].
\]
We also assume that the NBO (or NBB) is the superposition of two {\em independent} processes:
\[
    \NBO_t = p_t + \theta_t /2.0 = p_0 + \delta_t + \theta_t /2.0.
\]
The independence is a non-essential technical assumption, and one could otherwise work with the joint auto-covariance kernel for $(\delta_t, \theta_t)$ in later sections.

\subsection*{Instantaneous Cost： $\inst$}

Instantaneous cost represents the immediacy cost when execution consumes liquidity from the limit order books. In the reductionism approach when the market volume summarizes the overall market, market volume has to be somewhat proportional to the capacity or depth of the limit order book. This heuristic reasoning implies that bigger participation of volume (PoV) would efface deeper the limit order book and hence retreat the NBB more significantly (or NBO for sells). We therefore model this part of instantaneous cost (per share) by:
\[
    \inst_t = \alpha_1 \cdot h_t, \qquad \mbox{where $h_t = dX/dV$ is the PoV rate.}
\]
Here the coefficient $\alpha_1$ is in dollar and can be homogenized to a dimensionless one via:
\[
    \alpha_1 = \alpha_1^\circ \cdot p_0.
\]
Similar to the discussion for the spread cost, in practice $\alpha_1$ must  also depend on the underlying COP strategies and different execution houses may obtain different $\alpha_1$ when regressing against their own real trading data.

\subsection*{Transient Cost： $\tran$}

Transient cost represents the short-term market digestion of the most recent wave of trades. In the literature, this component is also  called the {\em resilience} term. For example, in the important work of Obizhaeva and Wang~\cite{algo_obizhaeva_wang13}, the authors attribute this term to the resilience of the limit order book.

What actually drives such a resilient force, however, has not been made very clear in the literature. Herein, we attribute it to the activities of quantitative market makers. A market maker adjusts her quoting levels based on the expected market behavior. As a common practice, the expected behavior is often based on the moving average of the observable. We assume that at any time $t$, a typical quantitative market maker is primarily interested in the expected PoV via the moving average of the observed PoV's:
\[
    \bar h_t = \int_0^t h_s dM(s | t),
\]
where $dM(s|t)$ is a backward moving average kernel. (Borrowed from probability theory, the vertical bar $(x | y)$ denotes
variable or value $x$ given $y$.) As a moving average kernel, one requires $dM(s|t)$ to be a probability measure over $s\in [0, t]$:
\[
    \int_{s \in [0, t]} dM(s | t) = 1.0, \myand 
    M(A | t) = \int_{s \in A} dM(s | t )  \ge 0,
\]
for any Borel set $A \subseteq [0, t]$. The most popular kernel used either in academia or in practical quantitative trading firms is the {\em exponential}  kernel (e.g.,~\cite{algo_obizhaeva_wang13}):
\[
    dM(s|t) = \frac 1 {Z_t} \exp( - \gamma (t- s) ), \qquad 0 \le s \le t.
\]
(The unital condition is often not strictly enforced, so that, for example, $Z_t \equiv  \gamma^{-1}$ or any suitable constant~\cite{algo_obizhaeva_wang13}.)

Based on the expected PoV $\bar h_t$,  the quantitative market maker then raises the quotes in an amount proportional to $\bar h_t$:
\[
    \tran_t = \alpha_2 \cdot \bar h_t = \alpha_2 \int_0^t h_s \cdot dM(s|t)
\]
Any incoming trade on $[t, t+dt)$ has to pay this premium cost.

In this work, we assume that $dM$ is {\em absolutely continuous} under the volume measure $dV$, and take the exponential decay kernel under volume distance as its Radon-Nikodym derivative:
\[
    dM(s | t) = \frac 1 {Z_t} \exp \left( -\frac {V_t - V_s} {V_\ast}   \right) dV_s,
\]
with the normalization constant $Z_t$ given by:
\[
        Z_t = \int_0^t \exp \left( -\frac {V_t - V_s} {V_\ast}   \right) dV_s= V_\ast \cdot \left(1 - \exp(-V_t/V_\ast) \right).
\]
Here the scaling constant $V_\ast$ reflects the market marker's volume-based {\em window} size for averaging short-term PoV's. For example, a market maker could take $V_\ast = 1\% \cdot ADV$, the {\em average daily volume} of the target security. (In the United States where the normal daily core session lasts 390 minutes, this is roughly the average trading volumes over several minutes.)

Since for the majority of time, $V_t \gg V_\ast$, $Z_t$ is quickly saturated to the level of  $V_\ast$. For the rest of the work we will make the following simplification by directly setting:
\[
    Z_t \equiv V_\ast.
\]
This constancy approximation is also the default setting for most existing works involving the {\em temporal} exponential kernel~\cite{algo_obizhaeva_wang13}. Analytically it helps alleviate the hardship involving a nonlinear exponential denominator.

\subsection*{Permanent Cost: $\perm$}

By definition, as time elapses and market volume grows, the transition cost imposed by a specific trade at some past time $s$ will fade away. The permanent cost component then captures any permanent impact a trade might have left. In the current work, we assume it is proportional to the {\em volume-weighted} average PoV:
\[
    \perm_t = \alpha_3 \int_0^t h_s \frac {dV_s}{V_t} = \alpha_3 \frac{ X_t} {V_t}.
\]
Most existing linear models only assume the proportionality to the total executed volume $X_t$ (e.g.,~\cite{algo_almgren_chriss00, algo_obizhaeva_wang13}). The volume normalization introduced herein captures the difference between the impact of 5000 shares, say, executed in a period of cumulative market volume $V_t = 20,000$ shares vs. that of the same number of shares in another period of $V_t = 200,000$ shares. The permanent impacts are clearly different. 

In theory the denominator $V_t$ vanishes in the beginning of the execution, as it is the cumulative market volume from the start at $t=0$.  This could potentially introduce some singularity when later on the total execution cost is assembled.  In the scenario of {\em flat} profiling when $V_t = v_0 t$ for some constant market rate $v_0$, for instance, the singularity is in the order of $O(1/t)$.  

In practice, quantitative traders  resolve such singularities through at least two general approaches. The first one is to slightly delay the computation in order to get stronger signals (with a higher {\em signal-to-noise ratio} (SNR) ). For instance,  many high-frequency strategies  on  the buy side  will not open to trade   within the first 5 to 30 minutes of the market opens, unless they specifically  target at the open auctions or opening moments. This ``quiet''  period allows to collect   {\em stable} signals based on moving averages.  The other approach is to {\em regularize} the target signals. In our scenario, for example, to ``boost'' $V_t$ to $V_t + \eps_0$ for some thresholding volume level $\eps_0$, so that the signal $V_t$ only becomes effective when $V_t  \gg \eps_0$. In practice, as discussed  for the transient cost,  $\eps_0$ could be $1\%\times ADV$, one minute average $ADV$,  or even a few or several round lots depending on the liquidity level of the security.  In the continuum setting, we use the $\eps$ symbol to convey a sense of minuteness, as common in mathematics. As such, we finalize the permanent cost term via:
\[
    \perm_t = \alpha_3 \int_0^t h_s \frac {dV_s}{V_t+\eps_0} = \alpha_3 \frac{ X_t} {V_t + \eps_0}.
\]
We point out that all the subsequent analysis also works for ``hard-thresholding'' when one takes:
\[
	\alpha_3 \int_0^t h_s \frac {dV_s}{V_t \vee \eps_0} = \alpha_3 \frac{ X_t} {V_t \vee  \eps_0} \qquad \mbox{instead,}
\]
where $V_t \vee \eps_0 = \max(V_t, \eps_0)$ more explicitly regularizes $V_t$ near $t=0$.

In actual discrete implementation, such regularization may not be necessary since $V_{t_1}$ is generally positive at the first grid point after $t=t_1 > 0$.

\subsection*{Putting Together}

Combining all the components, we arrive at a model for the {\em execution} price $\hat p_t$. Let $p_t$ denote the {\em market} price, which could be considered as the mid-quote of the NBBO at any time $t$. Then,
\begin{eqnarray}
    p_t         &=& p_0 + \delta_t + \tran_t + \perm_t; \\
    \hat p_t    &=& p_t + \sprd_t + \inst_t.
\end{eqnarray}
Here, $\delta_t$ represents the intrinsic market price movement, regardless of ``our'' own  trading activity. As discussed earlier, we only assume that its auto-covariance kernel $K_\delta(s, t)$ is known pre-trade.

\section{The Execution Model and Analysis}
\label{S:model_analysis}

The {\em implementation shortfall} (IS) is the amount overpaid (in the default scenario of buying) compared with the initial {\em paper} cost when the security price is $p_0$, as first introduced by Perold~\cite{algo_perold88}. In dollar amount, for any deterministic execution scheme : $X_t: 0 \le t \le 1$, it is defined by:
\[
\begin{split}
    \ISdlr  &= \int_0^1 (\hat p_t -p_0)\cdot dX_t \\
            &= \int_0^1 p_t \cdot dX_t + \int_0^1 \sprd_t \cdot dX_t + \int_0^1 \inst_t \cdot dX_t \\
            &=      \int_0^1 \delta_t \cdot dX_t + \int_0^1 \sprd_t \cdot dX_t + \int_0^1 \inst_t \cdot dX_t  \\
            &  \qquad \qquad \qquad + \int_0^1 \tran_t \cdot dX_t + \int_0^1 \perm_t \cdot dX_t \\
            &= \int_0^1 (\delta_t + \alpha_0 \theta_t\cdot \sign(h_t) ) \cdot dX_t \\
            & \qquad \qquad \qquad + \int_0^1 \left( \inst_t + \tran_t + \perm_t \right) \cdot dX_t.
\end{split}
\]
Due to the monotone condition discussed earlier, $\sign(h_t)$ is static and global since it is always 1 for a buy order  and $-1$ for a sell order.  Therefore,  we will substitute it with a static variable more commonly used:  $\side$, which is 1 for a buy and $-1$ for a sell. Notice that $\side \cdot dX_t = |dX_t|$. To prepare for the mean-variance formulation, we first compute the mean  $\E[ \ISdlr ]$, which is:
\begin{equation}		\label{E:ISdlr}
 \int_0^1 \left(  \alpha_0 \bar \theta_t \cdot \side + \alpha_1 h_t + \alpha_2 \int_0^t h_s dM(s|t) + \alpha_3 \int_0^t h_s \frac{ dV_s }{V_t+\eps_0}  \right)\cdot dX_t.
\end{equation}
The variance is
\begin{eqnarray} \label{E:ISdlr_var}
    \VAR[ \ISdlr ] &=& \int_0^1\int_0^1 K_{\delta,\theta}(s,t) dX_s dX_t, \qquad \mbox{with} \\
    K_{\delta,\theta}(s,t) &=& K_\delta(s, t) + \alpha_0^2 K_\theta(s, t).
\end{eqnarray}
For any given level of risk aversion expressed via a positive weighting parameter $\lambda > 0$, the mean-variance execution model is to find the optimal PoV function $\hat h_t$, such that $d\hat X_t = \hat h_t dV_t$ and the following utility functional is minimized:
\begin{equation}	 	
    J_\lambda[ h_t ] = \E[ \ISdlr ] + \lambda \cdot \VAR[ \ISdlr ].
\end{equation}

Under the given volume measure $dV_t$, the objective functional becomes:
\begin{equation} 	\label{E:objective:mean_variance}	
\begin{split}
    J_\lambda[ h_t ] & = \alpha_0 \cdot \side \cdot \int_0^1 \bar \theta_t h_t dV_t + \alpha_1 \int_0^1 h_t^2 dV_t \\
                     & + \frac{\alpha_2}{2V_\ast} \int_0^1 \int_0^1 h_s h_t \exp\left( -\frac{|V_t - V_s|}{V_\ast}\right) dV_s dV_t \\
                     & + \frac{\alpha_3} 2 \int_0^1 \int_0^1 h_s h_t \frac{1}{V_t \vee V_s + \eps_0} dV_s dV_t \\
                     & + \lambda \int_0^1 \int_0^1 K_{\delta,\theta}(s, t) h_s h_t dV_s dV_t,
\end{split}
\end{equation}
with at least two common constraints discussed in previous sections,
\begin{equation} 	\label{E:constraints:monotone_complete}
    \side \cdot h_t \ge 0, \myand \int_0^1 h_t dV_t = X_1,
\end{equation}
for monotonicity and completion. Notice also that $\side = \sign(X_1)$. 

According to the published marketing or sales sheets,   major execution houses always allow their clients to specify the PoV capping, i.e., the linear inequality constraint:
\begin{equation}		\label{E:constraints:volumelimit}
		\side \cdot h_t \le \mathrm{maxPoV}. 
\end{equation} 		
For most clients, the popular comfort zone for maxPoV is somewhere between $5\%$ and $25\%$.  If this constraint is turned on, there is an obvious compatibility condition for the model to yield a  solution: 
\begin{equation} 	\label{E:constraints:compatible}
	 \int_0^1 \maxPoV \cdot dV_t \ge \side \cdot X_1, \myor \maxPoV \ge \frac {|X_1|}{ V_1}.
\end{equation}
Otherwise, the execution may not be complete within the specified horizon.  Except for very urgent or small orders, clients usually turn on this constraint as an ultimate safeguard.

%%% 002.M0610.2013
\subsection{Basic Assumptions on the Volume Measure}

Since both the transient and permanent cost terms explicitly involve the cumulative market volume function:
\[
	V_t =  \int_{[0, t]} dV_s,
\]
we will make the following assumptions about the regularity of the volume measure.

\begin{enumerate}[(i)]
\item (Absolute Continuity) We assume that $dV$ is {\em absolutely continuous} with respect to the ordinary   Lebesgue measure $dt$. This amounts to saying that there exists a non-negative function $v_t \in L^1(I) = L^1(I, dt)$, such that $dV_t = v_t dt$. Then the cumulative market volume function
\[ V_t = \int_{[0,t]} v_t \cdot dt, \]
is well defined {\em pointwise}, and is continuous and non-decreasing.  We call $v_t$ the {\em market (trading) rate} function. 

\item (Lipschitz) Furthermore, we assume that the market rate function is bounded from above, i.e., there exists some constant $A>0$, such that 
 \[
 	v_t \le A, \qquad \mbox{almost everywhere $t \in I=[0, 1]$ under $dt$.}
 \]
 Equivalently, we say that $v_t \in L^\infty(I)$. In terms of the cumulative market, it is equivalent to requiring $V_t$ to be Lipschitz~\cite{evagar,fol}.
 
\end{enumerate} 

Practically these are natural assumptions to most execution houses, where volume profiles are generated by overnight processes. One common component of these processes is a smoothing kernel to curb the effect of spurious spikes arising from direct historical averaging.

\subsection{Convexity and Uniqueness}

Define the kernel functions:
\[
	K_2(s, t) =\frac 1 {V_\ast} \exp\left( -\frac{|V_t - V_s|}{V_\ast}\right), \myand
	K_3(s, t) = \frac 1 {V_t \vee V_s + \eps_0}. 
\]
Then the objective functional $J_\lambda[h]$ involves three quadratic terms in the general form of
\[
	Q_{K, dV}[h] = \int\int_{I \times I} K(s, t) h(s) h(t) dV(s) dV(t),
\]
with $K=K_2, K_3,$ or $K_{\delta,\theta}$ and $I=[0,1]$.  In what follows, we assume that the kernel function is symmetric: $K(s,t)= K(t,s)$ and that
\[
	K: I \times I \to \mathcal{R} \qquad \mbox{is continuous}. 
\]
Here one notices the regularization role of $\eps_0$ introduced for the permanent cost component,  without which $K_3$ would not be continuous at $(t,s)=(0,0)$.  The following  norm formula is also standard in functional analysis~\cite{evagar, fol}
\[
	\|Q_{K,dV}\| := \sup_{\|h\|_{L^2(dV)} \le 1} Q_{K,dV}[h] 
			\le \| K\|_{L^2(dV \otimes dV) } 
			\le \|K \|_{L^\infty} \cdot V_1. 
\]

\begin{definition} 
$Q_{K,dV}[h]$ is said to be {\em positive} in the Hilbert space $L^2(I, dV)$ if for any $h \in L^2(I, dV)$, $Q_{K,dV}[h] \ge 0$. A symmetric and continuous kernel $K(s, t)$ on $I \times I$ is said to be {\em positive}, if for any finite sequence $(t_i | i=1:N)$ in $I$ of arbitrary length $N$,  and real scalars $(c_i | i=1:N)$, one has
\[
	\sum_{i,j=1}^N K( t_i, t_j ) c_i c_j \ge 0. 
\]

\end{definition}

Notice that in some literature, it is said to be {\em nonnegative}. 

\begin{proposition}
The quadratic form $Q_{K,dV}[h]$ induced by a positive kernel $K$ must be positive.
\end{proposition}

\beginProof
This is canonical in the context of {\em ordinary} Lebesgue measure $dt$, which we do refresh here first. Assume that $K$ is a positive kernel. First notice that
\[
	Q_{K,dt}[g] =\int\int_{I \times I} K(s, t) g(s) g(t) ds dt 
				\le \|K\|_{L^\infty(dt)} \|g\|^2,
\]
for any $g \in L^2(I, dt)$. Thus $Q_{K,dt}$ is positive if and only if it is positive on a dense set of $L^2(I, dt)$. It is well known in real analysis~\cite{fol} that the set of continuous functions $C(I)$ is indeed dense in $L^2(I, dt)$. For any $\phi(t) \in C(I)$, the   Lebesgue integral in $L(I\times I, dt \otimes ds)$
\[  
	Q_{K,dt}[\phi] = \int\int_{I \times I} K(s, t) \phi(s) \phi(t) ds dt 
\]
is identical to its Riemann integral, since $K(s,t)\phi(s)\phi(t) \in C(I \times I)$. For any partition of $I$:
\[
	0=t_0 < t_1 < t_2 < \cdots < t_N = 1,
\]
the following Riemann sum is positive since $K$ is assumed positive:
\[
	\sum_{i,j=1}^N K( t_i, t_j ) \phi(t_i) \phi(t_j) (t_{i+1}-t_i)(t_{j+1}-t_j) \ge 0.
\]
Taking proper limits, one sees that $Q_{K,dt}$ is indeed positive on $C(I)$.

Now consider a general volume measure $dV$ with $v = dV/dt \in L^\infty(dt)$. For any $h_t \in L^2(dV)$, one has
\[ 
	\int_I h^2 v^2 dt \le \|v\|_\infty \cdot \int h^2 vdt =\|v\|_\infty \|h\|^2_{L^2(dV)}, 
\]
implying that $g(t)=h(t) v(t) \in L^2(dt)$. Then the proof is complete following that 
\[
	Q_{K, dV}[h] = Q_{K,dt}[g] \ge 0.  
\]
\closeProof

\begin{proposition}
The risk kernel $K= K_{\delta, \theta}$ is positive.
\end{proposition}

By definition, $K_{\delta, \theta}(t, s)$ is the auto-covariance function of the price-spread mixed process:
\[ 
	W_t = \delta_t + \alpha_0 (\theta_t - \bar \theta_t),
	\myand K_{\delta, \theta}(t, s) = \E[ W_t \cdot W_s ]. 
\]
Then for any finite sequence $(t_i|i=1:N)$ and real scalars $(c_i | i=1:N)$, 
\[
	\sum_{i,j=1}^N K_{\delta, \theta}(t_i, t_j) c_i c_j 
		= E\left[ \left(\sum_{i=1}^N c_i W_{t_i}    \right)^2 \right] \ge 0,
\]
which establishes the positivity of $K_{\delta, \theta}$. 

To proceed further, we need a useful lemma whose proof follows directly from the definition of kernel positivity. 
\begin{lemma}
Suppose $K(T, S)$ is a positive kernel on a subset $ D \subseteq \realR $, and $\phi: I \to D$ any real function with $T=\phi(t)$. Then the {\em pullback} kernel $K_\phi(t, s) = K(\phi(t), \phi(s))$ is positive on $I$. 
\end{lemma}

\begin{proposition}
The transient kernel $K=K_2$ is positive. 
\end{proposition}

\beginProof
Given a real function $g(t)$ on $t \in \realR$, suppose its Fourier transform $\ds G(\omega) = \int_\realR g(t) e^{ -\sqrt{-1} t\omega} dt$ is real and non-negative for all $\omega \in \realR$. Then the kernel defined via:
\( \ds K_g(t, s) = g(t-s) \)
must be positive for $t, s \in \realR$. This is because:
\[
\begin{split}
	\sum_{i,j=1}^N K_g(t_i, t_j) c_i c_j
		&= \sum_{i,j=1}^N g(t_i - t_j) c_i c_j \\ 
		&= \sum_{i,j=1}^N \frac 1 {2 \pi} \int_\realR G(\omega) e^{ -\sqrt{-1} (t_i-t_j) \omega} d\omega \cdot c_i c_j \\
		&= \frac 1 {2\pi} \int_\realR G(\omega) 
		\big| \sum_{i=1}^N c_i e^{ -\sqrt{-1} t_i \omega}  \big|^2 d\omega \ge 0.
\end{split}
\]
Since the Fourier transform of an exponential $g(T) = \gamma \exp(-\gamma|T|)$ with $T\in \realR$ is 
\( \ds
	G(\omega) = \frac { 2\gamma^2 } { \gamma^2 + \omega^2 } ,
\)
we  conclude that 
\[
	K_g(T, S) = g(T- S) = \gamma \exp(-\gamma|T - S|)
\]
must be positive on $T, S \in  \realR$.  The proof is then complete by taking $\gamma= 1/V_\ast$, and $\phi(t)=V_t$ in the preceding lemma. 
\closeProof

\begin{proposition}
The permanent kernel $K=K_3(t, s)$ is positive.
\end{proposition}

\beginProof
First it is easy to see from the definition that, if $K(T, S)$ is positive on a subset $D \subseteq \realR$, and $f(t): D \to \realR$ a real function,  the new  kernel
\[	
	K_f(T, S) = K(T, S) f(T) f(S)
\]
must be positive on the same domain.  Now define $D=(0, \infty)$, and 
\[  K_B(T, S) = T \wedge S, \myand f(T) = 1/T.  \]
Notice that $K_B(T, S)$ is positive since it is the auto-covariance of the canonical Brownian motion $B_T$:
\[
	K_B(T, S) = \E\left[ B_T \cdot B_S  \right],  \qquad T, S > 0.
\]
Therefore, the kernel 
\[
	K(T, S) = \frac 1 {T \vee S} = \frac {T \wedge S}{T \cdot S} = K_B(T, S) f(T) f(S)
\]
must be positive on $D$. The proof is complete via the preceding lemma with $\phi(t) =\eps_0+ V_t$. 
\closeProof

\begin{theorem}[Convexity and Uniqueness]  The objective functional $J_\lambda[h_t]$ is {\em strictly} convex in $L^2(I, dV)$. As a result,  the optimal execution solution $h^\ast_t$ to the constrained model must be {\em unique}.
\end{theorem}

\beginProof
The preceding propositions establish that the last three quadratic forms in $J_\lambda[h]$ are all positive and thus convex in the Hilbert space $L^2( dV)$. The second quadratic term (from instantaneous cost)
\[
	\alpha_1 \int_I h_t^2 dV_t = \alpha_1 \|  h_t \|^2 _{L^2(dV_t)}
\] 
is the squared norm and hence strictly convex. Since the first term on the average spread cost is linear, $J_\lambda[h]$ must be {\em strictly} convex. 
\closeProof

\subsection{Compactness and Existence}

For convenience, define the combined kernel
\begin{equation}	 		\label{E:operatorK}
	\begin{split}
		\opK(t, s) 
		&=  \frac{\alpha_2} 2 K_2(t, s)  + \frac{\alpha_3}{2} K_3(t, s) + \lambda K_{\delta, \theta}(t, s) \\
		&= 	\frac{\alpha_2}{2 V_\ast}\exp\left(  - \frac {|V_t-V_s|} {V_\ast}  \right) 
				+\frac {\alpha_3} 2 \frac 1 { V_t \vee V_s + \eps_0 } 
				+ \lambda K_{\delta, \theta}(t, s).
	\end{split}
\end{equation}
Since $\opK(t, s)$ is a continuous kernel over $I \times I = [0, 1]^2$,  we have 
\[
	 \opK(t, s)  \in L^2(I \times I, dV \otimes dV),  
	 \;\text{and}\;  
	 \| \opK(t, s) \|_{L^2(dV \otimes dV} \le \|\opK(t, s) \|_\infty \cdot V_1. 
\]
We also use the same symbol to denote the induced linear operator in $L^2( I, dV)$:
\[
	\opK h_t := \int_I \opK(t, s) h(s) dV_s. 
\]
It is well known~\cite{book_analysis_eidmiltso2004} that (i) the $L^2$ {\em  function} norm dominates the {\em operator} norm:
\[
	\| \opK\|  \le \| \opK \|_{L^2(dV \otimes dV)} < \infty,
\]
and (ii) such a linear operator $\opK$ must be {\em compact} in the Hilbert space of $L^2(I, dV)$. 

\begin{theorem}
There exists a unique optimal execution solution $h^\ast_t$ to the mean-variance model with $J_\lambda[h]$ defined in Eqn.~(\ref{E:objective:mean_variance})   as the objective function, and with  the following constraints (Eqn.~(\ref{E:constraints:monotone_complete}) and (\ref{E:constraints:volumelimit})): monotonicity, completion, and PoV capping, as long as the latter two are compatible as expressed in Eqn.~(\ref{E:constraints:compatible}).  
\end{theorem}

\beginProof
By Theorem~1, it suffices to further establish the existence portion.  Define
\[
	\hat \phi_t = \alpha_0 \cdot \side \cdot \bar \theta_t. 
\]
Then the objective $J_\lambda[\cdot]$ can be expressed more  compactly by:
\[
	J_\lambda[h] = \inner{\hat \phi_t}{h_t} + \alpha_1 \inner{h_t}{h_t}  + \inner{\opK h_t}{h_t},
\]
where the inner product is in the Hilbert space of $L^2(I, dV)$. Let $\opI$ denote the identity operator. Then we have
\[
	J_\lambda[h] = \inner{\hat \phi_t}{h_t} + \inner{ (\alpha_1 \opI + \opK)h_t }{ h_t }.
\]
Since $\opK$ is the linear combination of three positive operators (or kernels) with positive coefficients, it must be {\em positive}.  It is well known in the {\em spectral theory}~\cite{book_analysis_eidmiltso2004}   that the spectrum set $\sigma(\opK)$ of such a compact and positive operator must be a subset of the positive half real axis $[0, \infty)$, and for any $\mu \notin \sigma(\opK)$,  the inverse $\opK - \mu \opI$ must exist and be bounded in the Hilbert space $L^2(I, dV)$.  In particular, with $\mu = - \alpha_1 \notin \sigma(\opK)$,  $(\opK + \alpha_1 \opI)^{-1}$ is a well-defined bounded linear operator, and one can define
\[
	\phi_t = (\alpha_1 \opI + \opK)^{-1} \hat \phi_t \in L^2(I, dV).
\]
We also further introduce a new symmetric bilinear function: for $g, h \in L^2(I, dV)$, 
\[
	\outer{g}{h} = \inner{ (\alpha_1 \opI + \opK)g_t}{ h_t}
\]
It is {\em strictly} positive since it {\em dominates} the ordinary inner product:
\[
	\outer{h}{h} \ge \alpha_1 \inner{h}{h}. 
\]
Thus it introduces a new inner product, which is actually {\em equivalent} to the natural one since  it is also bounded above:
\[ 
	 \outer{h}{h}  = \alpha_1 \inner{h}{h} + \inner{\opK h}{h} \le (\alpha_1 + \|\opK\|_\infty \cdot V_1) \inner{h}{h}. 
\]
For convenience, we denote by $L^2(I, dV | (\cdot, \cdot) )$ the same function space $L^2(I, dV)$ endowed with this new equivalent inner product, which is a Hilbert space.  Furthermore,  the original objective function $J_\lambda[h]$ simplifies to:
\begin{equation} \label{E:objective:newform}
	J_\lambda[h] = \outer{\phi_t}{h_t} + \outer{h_t}{h_t}. 
\end{equation}
Define $\psi_t = (\alpha_1 \opI + \opK)^{-1} 1$.  Then the completion constraint becomes:
\(
	\outer{\psi_t}{h_t} = X_1. 
\)
And the constraints on monotonicity and participation limit remain the same:
\[
			0 \le \side \cdot h_t \le \maxPoV.
\]
The constant PoV execution strategy:
\[ 
		h_t^{const} \equiv \frac {X_1}{V_1}, \qquad t \in I=[0,1]
\]
is clearly admissible under the given constraints due to the compatibility assumption in Eqn.(~\ref{E:constraints:compatible}), and has a  finite objective value. Then there must exist a non-empty sequence of minimizing execution strategies with finite objectives $J_\lambda$'s:
\[
		h^{(1)}_t, h^{(2)}_t, \cdots, 
\]
in $L^2(I, dV)$, such that 
\[   \lim_{n \to \infty} J_\lambda[h^{(n)}] =   \inf J_\lambda[h] < \infty, \]
and each meets the constraints.  The sequence must be bounded since one can easily show from Eqn.~(\ref{E:objective:newform}) that
\[
	\| h_t  \|_{(\cdot, \cdot)} 
			 \le \left(  J_\lambda[h_t] + \frac 1 4  \| \phi_t \|^2_{(\cdot, \cdot)}  \right)^{1/2}
		 + \frac 1 2 \| \phi_t \|_{(\cdot, \cdot)}.
\]
Now that in Hilbert spaces, any bounded sequence must be {\em weekly} pre-compact, possibly replaced by one of its subsequences, $(h^{(n)}_t | n=1, 2, \cdots)$ could be assumed to {\em weakly} converge to some element $h^\ast_t \in L^2(I, dV | (\cdot, \cdot) )$, so that for any $g \in L^2(I, dV | (\cdot, \cdot) )$:
\[
		\outer{g_t}{h^\ast_t} = \lim_{n \to \infty} \outer{g_t}{h^{(n)}_t}. 
 \]
 Since the Hilbert norm is known to be {\em lower semi-continuous} (l.s.c.) under week convergence, we have 
 \[
	J_\lambda[h^\ast] \le  \lim_{n \to \infty} \outer{\phi_t}{h^{(n)}_t}  
			+ \liminf_{n \to \infty} \outer{ h^{(n)}_t }{ h^{(n)}_t } =  \liminf_{n \to \infty} J_\lambda[h^{(n)}_t] 
			= \inf J_\lambda[h]. 
 \]
 Finally the admissible space defined by the three constraints is easily seen to be a {\em closed} and  {\em convex set}, which must be {\em closed} under weak convergence~\cite{book_analysis_eidmiltso2004}. This means $h^\ast$ also meets all the three constraints on completion, monotonicity, and PoV capping. Then $h^\ast$ must be the optimal execution strategy with $J_\lambda[h^\ast] = \inf J_\lambda[h]$. 
\closeProof

\section{Computation and Numerical Examples}	
\label{S:computation}

\subsection{Model Calibration}    \label{secsub:calibration}

In this subsection, we briefly explain the major steps for model calibration. Actual implementation should depend on the trade database each execution house owns.  For example, as explained in Section~\ref{S:costcomponents}, COP strategies employed by an execution house may directly impact  data bookkeeping and subsequent model calibration, 

Although the proposed model involves a varieties of kernels, its main characteristics is the linearity  for all  its four major  parameters: $\alpha_0, \alpha_1, \alpha_2$, and $\alpha_3$.  Model calibration can thus be directly based upon linear regression.

We first make the following assumptions about the empirical trades already made in the past and stored in the database of the execution house. 

\begin{enumerate}[(i)]
\item  (Universe Coverage)  For targeted securities, one has enough samples of past trades. For any major broker-dealers, exchanges, or agency houses, this is typically not an issue. For example, in the United States, securities from the universes of all major indices (e.g.,  S\&P 500) are heavily traded.  (Due to the proprietary nature of trade information, this often imposes much challenge for academic researchers.)

\item (Data Cleaning) Trade data have been properly {\em filtered}.  This has been a very common practice in execution houses. To filter is to remove erroneous trades or insignificant trades from participating in the calibration. For example, trades lasting less than 5 minutes or trades whose average PoVs are below 0.1\% can be considered insignificant and filtered out.

\item (Recording Intervals) For each trade, the house has kept on record its execution details, which may include for example, shares traded over each 10 seconds and average execution prices associated with. The  periodic recording interval should not be too long so that the calibrated model can properly capture the transient effect. 

\item (Profiles and Risks) Using statistical methods and consolidated exchange data, the house has already created the standard profiles for the Bid-Ask spread $\bar \theta_t$, and the volume measure $dV_t$  for each security, as well as the risk metrics for the auto-covariance matrices $K_\delta(t, s)$ and $K_\theta(t, s)$ as risk metrics.   
\end{enumerate}

At the execution houses where the author had worked, trading models are typically calibrated daily, or weekly the longest. (Risk parameters involved could stay longer, however.)

Implementation shortfalls (IS) are typically recorded and reported as basis points (bps). One basis point is $0.0001$ (of a given currency). We thus define $\bps = 1/10,000$.  For a given trade, IS in bps
is related to IS in dollars (or any other working currency) by
\begin{equation}  \label{E:ISbps}
	\ISbps = \frac \ISdlr {p_0 |X_1|} \cdot  \frac 1 {\bps}. 
\end{equation}
For example, if a buying trade of targeted initial notional $p_0 * |X_1| = \$10,000,000$ is reported to have 12 bps of $\ISbps$, it means the final net cost the client actually has to pay is $ \$10,012,000$. This even does not include any trading commissions or fees. 

In order to calibrate the model to the reported $\ISbps$  in databases, and to have normalized magnitudes for the model coefficients, we make the following coefficient normalization based on proper dimensionality analysis (with the circular superscript $\square^\circ$ representing normalized coefficients)
\begin{align*}
	\bar \theta_t  & = \bar \theta_t^\circ \cdot  p_0 \cdot \bps, \qquad 	\alpha_0 = \alpha_0^\circ \cdot \\
	\alpha_i & = \alpha_i^\circ \cdot p_0 \cdot \bps, \qquad i=1, 2, 3.
\end{align*}
Then by Eqn.~(\ref{E:ISdlr}) for $\ISdlr$, and the definition for $\ISbps$  in Eqn.~(\ref{E:ISbps}), 
\begin{align*}
\E\left[ \ISbps[h] \right]
	&= 		 \alpha^\circ_0 \int_I \bar  \theta^\circ_t \frac{dX_t}{X_1}
			+ \alpha^\circ_1 \int_I h_t  \frac{d X_t}{|X_1|} \\
	&\qquad\qquad		
	 + \alpha^\circ_2 \int_I   \frac{dX_t}{|X_1|} \int_0^t h_s dM(s | t)
	 + \alpha^\circ_3  \int_I   \frac{dX_t}{|X_1|} \int_0^t h_s \frac{dV_s}{V_t +\eps_0}  \\
	&=		\alpha^\circ_0 \cdot C_0[h] + \alpha^\circ_1 \cdot C_1[h] 
			+ \alpha^\circ_2 \cdot C_2[h] + \alpha^\circ_3 \cdot C_3[h] 
\end{align*}
We thus apply linear regression to all validated historical trades (i.e., $h$'s already observed) based on the linear model:
\[
	\ISbps[h] \sim \alpha^\circ_0 \cdot C_0[h] + \alpha^\circ_1 \cdot C_1[h] 
			+ \alpha^\circ_2 \cdot C_2[h] + \alpha^\circ_3 \cdot C_3[h]  + w[h].
\]
Notice the {\em heteroskedasticity} of the model as the residual noise term $w[h]$  carries a variance that depends on the execution profile $h_t$, as shown in Eqn.~(\ref{E:ISdlr_var}). We thus apply either the weighted least-square estimator (w-L.S.E.) or the equivalent ordinary L.S.E. to the variance normalized data. 

We leave some other auxiliary details to the execution houses who are interested in the current work and who can conduct actual model calibration from their proprietary trading database. The author welcomes any feedback or suggestion from the industry. 

\subsection{Numerical Computing via Quadratic Programming}

In this section, we show how the proposed model can be efficiently computed via quadratic programming algorithms and available commercial software (e.g., MOSEK or IBM CPLEX). 

Unlike dynamic trading models adapted to evolving market environments, pre-trade models are typically built upon {\em historical} ``profiles'' for spreads, volumes, correlations and volatilities. These profiles are often generated via robust statistical methods daily or weekly, and stored into data files. Typically they are discrete intraday vectors with intervals ranging from 30 seconds to 5 minutes, based on both the liquidity levels of the target securities  and the COP machinery each house employs. Therefore, below we study the discrete implementation on a regular discrete time grid. 

Throughout the work we have been using the normalized trading horizon $I=[0, 1]$, mainly to introduce both notional and analytical convenience and clarity. This is not necessary for actual computing and we can work directly with the physical time $I=[T_0, T_1]$, say, from 9:45am to 1:35pm.  Given a time interval $\Delta t$, suppose the trading horizon is discretized to:
\[  t_0 = T_0, t_1 = \Delta t, \cdots, t_N = T_1. \]
Then the continuous volume measure $dV$ is discretized to a discrete volume ``profile'':
\[
		d_n = V\left( [t_{n-1}, t_n) \right)  = \int_{t_{n-1}}^{t_n} dV_t, \qquad n=1:N,
\]
which is usually made available by the analytics teams of execution houses.          Let 
\[ \opD = \mathrm{diag}(d_1, \cdots, d_N)\]
be the diagonal matrix. The target PoV function $h_t$ is similarly discretized to a column vector $H=(h_1, \cdots, h_N)^T$, with
\[
	h_n = \frac 1 {d_n} \int_{t_{n-1}}^{t_n} h_t dV_t, \qquad n=1:N,
\]
whenever $d_n > 0$ (i.e., market volume is non-zero), and $h_n=0$ otherwise. 

Following the operator expression in Eqn.~(\ref{E:operatorK}), we recycle the same symbol of $\opK$ to denote the matrix corresponding to the discretization of the continuous kernel $\opK(t, s)$.  Let $B=(b_1, \cdots, b_N)^T$ denote the column vector with
\[
	b_n = \alpha_0 \cdot \side \cdot \bar \theta_{t_{n-1/2}}, \quad \mbox{with} \; 
	{t_{n-1/2}} = \frac { t_{n-1} + t_n } 2, \qquad n=1:N. 
\]
Then the model is discretized to:
\begin{align*}
\mbox{minimize over $H$: }  	\quad 
						& B^T\cdot \opD \cdot H + H^T \cdot  (\alpha_1 \opD + \opD \cdot \opK \cdot \opD)  \cdot H \\
\mbox{subject to: } \quad   &   \\
 					  &	\side \cdot H \le \maxPoV; \\
 					  &	\mathrm{diag}(\opD)^T \cdot H = X_1; \\
 					  &	\side \cdot h_n \ge 0, \;\;  \mbox{where $d_n >0$ }, \qquad \mbox{and}  \\
 					  &	h_n \equiv 0, \;\;  \mbox{where $d_n  = 0$}.
\end{align*}
Here $\mathrm{diag}(\opD)$ denotes the column vector consisting of the diagonals of $\opD$ (following MATLAB).
The last condition can also be used to reduce the actual dimension of the problem by eliminating
zero PoV's where there are no market volumes (i.e., with $d_n=0$).  

This discrete problem fits well into the framework of quadratic programming, and can be efficiently solved numerically by commercial optimizers such as MOSEK and IBM CPLEX, which are often integrated into the local Java or C++ libraries of in-house execution analytics.

\subsection{Numerical Examples}

In this section, we present several  examples that help reveal  the general behavior of the proposed model. We have relied on  the built-in quadratic programming optimizer {\em quadprog.m} in MATLAB.   The MATLAB software generating these examples are available from the author upon request.

Throughout we assume a hypothetical market that opens for 390 minutes from 9:30am to 4:00pm. The target security is assumed to have an arrival price of $p_0 = \$30.00$, and be moderately liquid with {\em average daily volume} (ADV) about 5,000,000 shares. We also assume that the volume and spread profiles have been established discretely in minutes, and that the volume profile bears a typical U-shape with more volumes at  the Open and  Close.  The primary task is to buy  $X_1 = 90,000$ shares during  a horizon that spans 90 minutes.  

Under these general settings, we assume that the hypothetical client is allowed to customize on the following three trading factors: (1) starting time $T_0$,  (2) PoV capping $\maxPoV$, and (3) level of risk aversion $\lambda$.  The choice of starting time affects the ``shape'' of the volume measure over the trading horizon due to the daily U-shape.  In most examples we set $\maxPoV = 20\%$, and the  risk aversion to a medium level of $\lambda=10^{-3}$.  To better compare with the existing literature, unless indicated otherwise,  the price dynamics is assumed to be Brownian with constant spreads.

\subsubsection{Effect of Risk Aversion}

\begin{figure}[h]
\centering
\includegraphics[width=0.75\textwidth, height=3in]{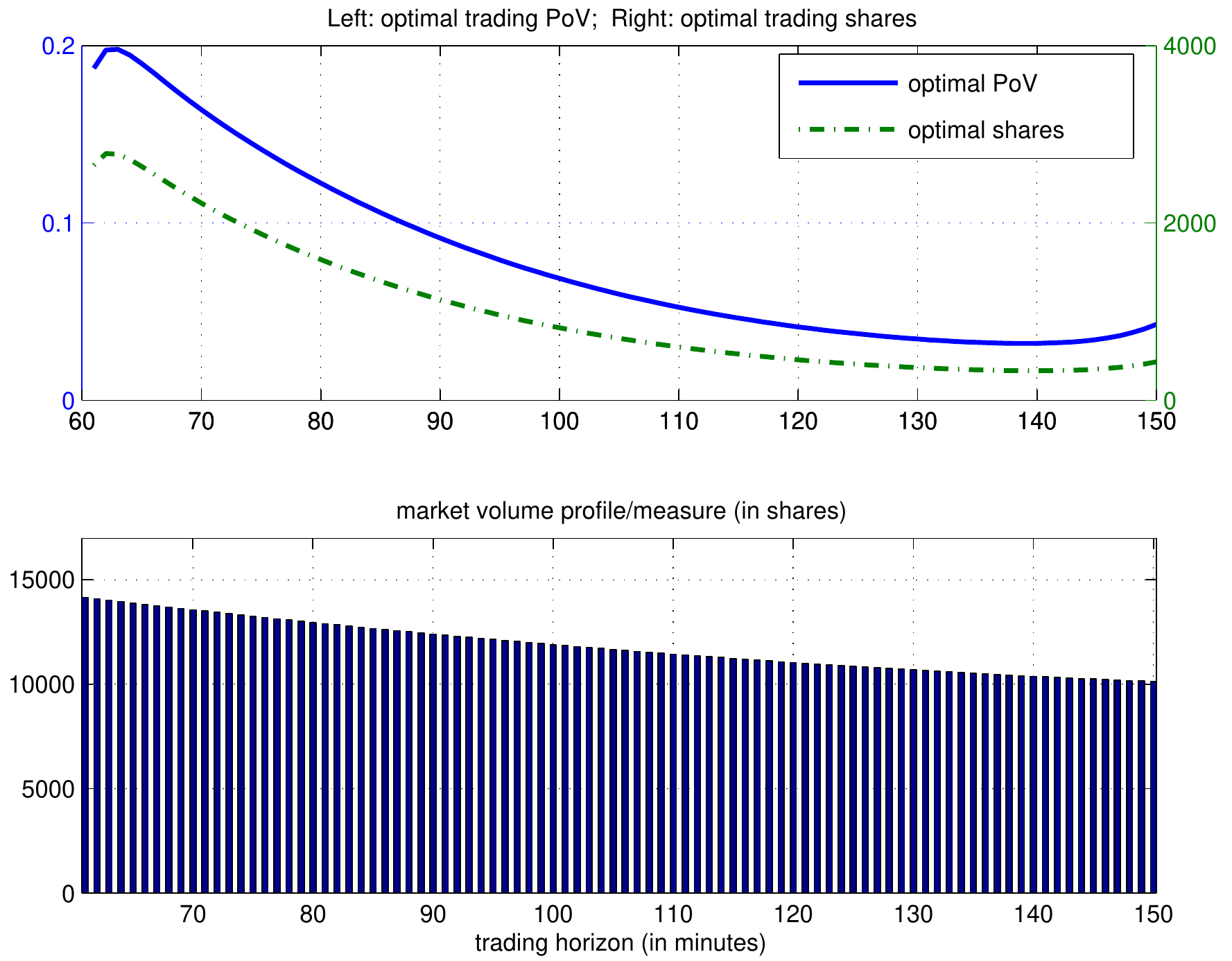}
\caption{Medium risk aversion with $\lambda=10^{-3}$ }
\label{F:RA_Medium}
\end{figure}

\begin{figure}[h]
\centering
\includegraphics[width=0.51\textwidth]{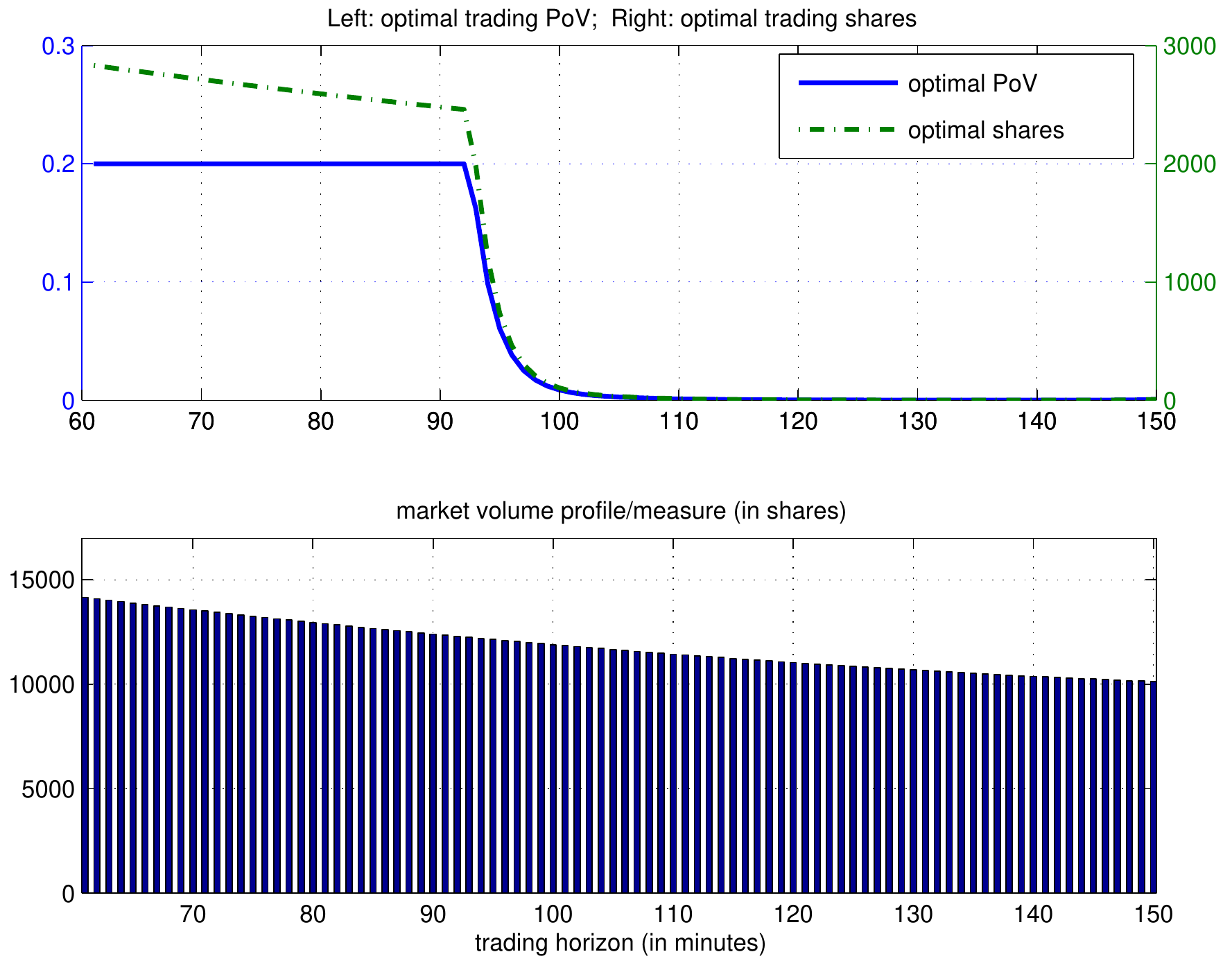} 
\nolinebreak
\includegraphics[width=0.51\textwidth]{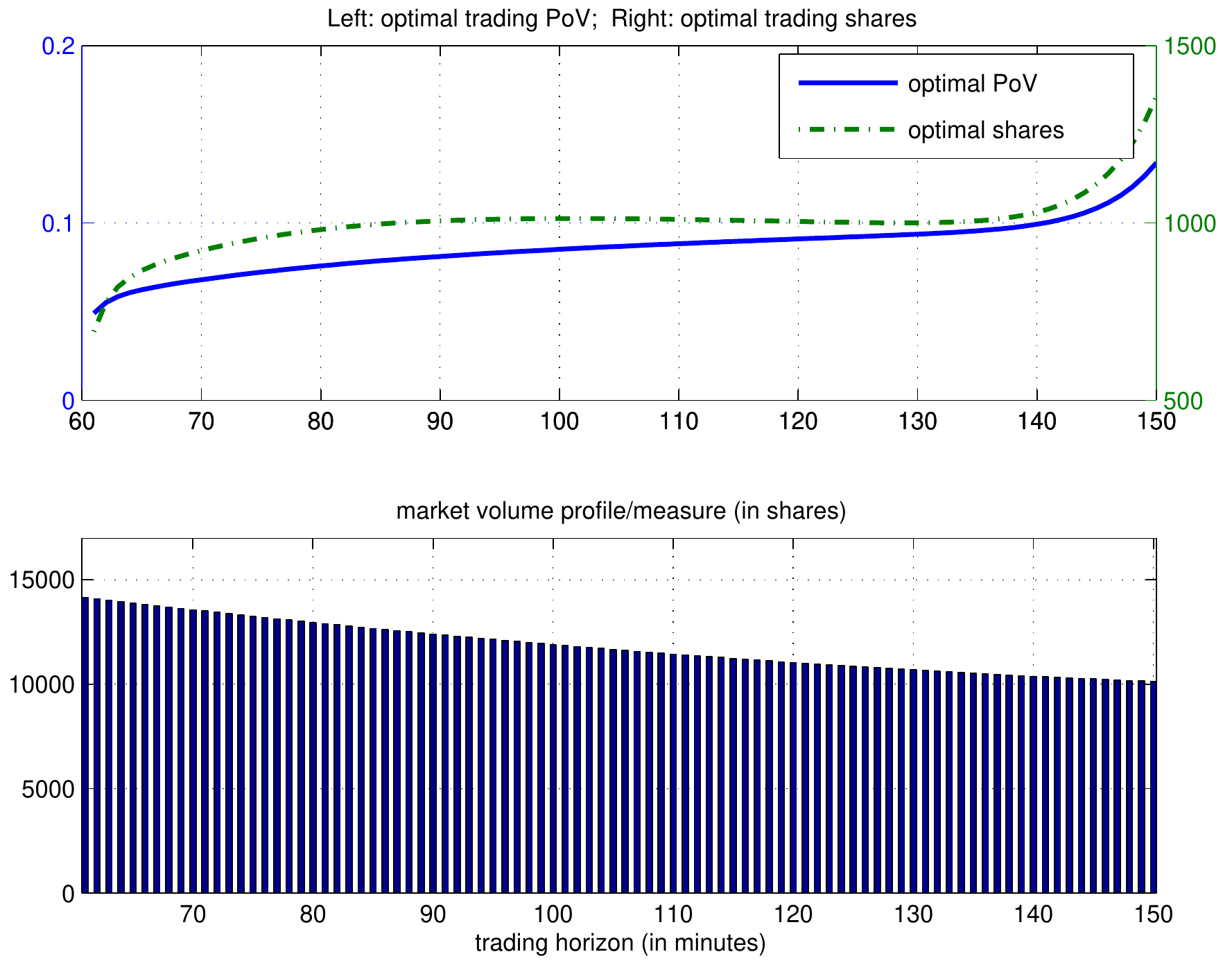}
\caption{ (Effect of Risk Aversion) Left: high risk aversion with $\lambda=10^{-1}$; Right: low with $\lambda=10^{-5}$. Higher risk aversion implies more front-loading behavior. Unlike many results in existence, front loading is allowable only up to the level of the $\maxPoV$  set by a client. }
\label{F:RA_HighLow}
\end{figure}

Plotted in Fig.~\ref{F:RA_Medium} and Fig.~\ref{F:RA_HighLow} are the optimal execution solutions corresponding to three different levels of risk aversion: medium (Fig.~\ref{F:RA_Medium}), and high and low (left and right panels in Fig.~\ref{F:RA_HighLow}).
As common for mean-variance models, more risk aversion implies more front-loading behavior. But unlike most results illustrated in the classical literature, front-loading is allowable only up to the level of the $\maxPoV$ ( which is 20\% in this series of examples) typically set by the clients. 

\subsubsection{ Effect of Volume Measures }

\begin{figure}[h]
\centering
\includegraphics[width=0.75\textwidth, height=3in]{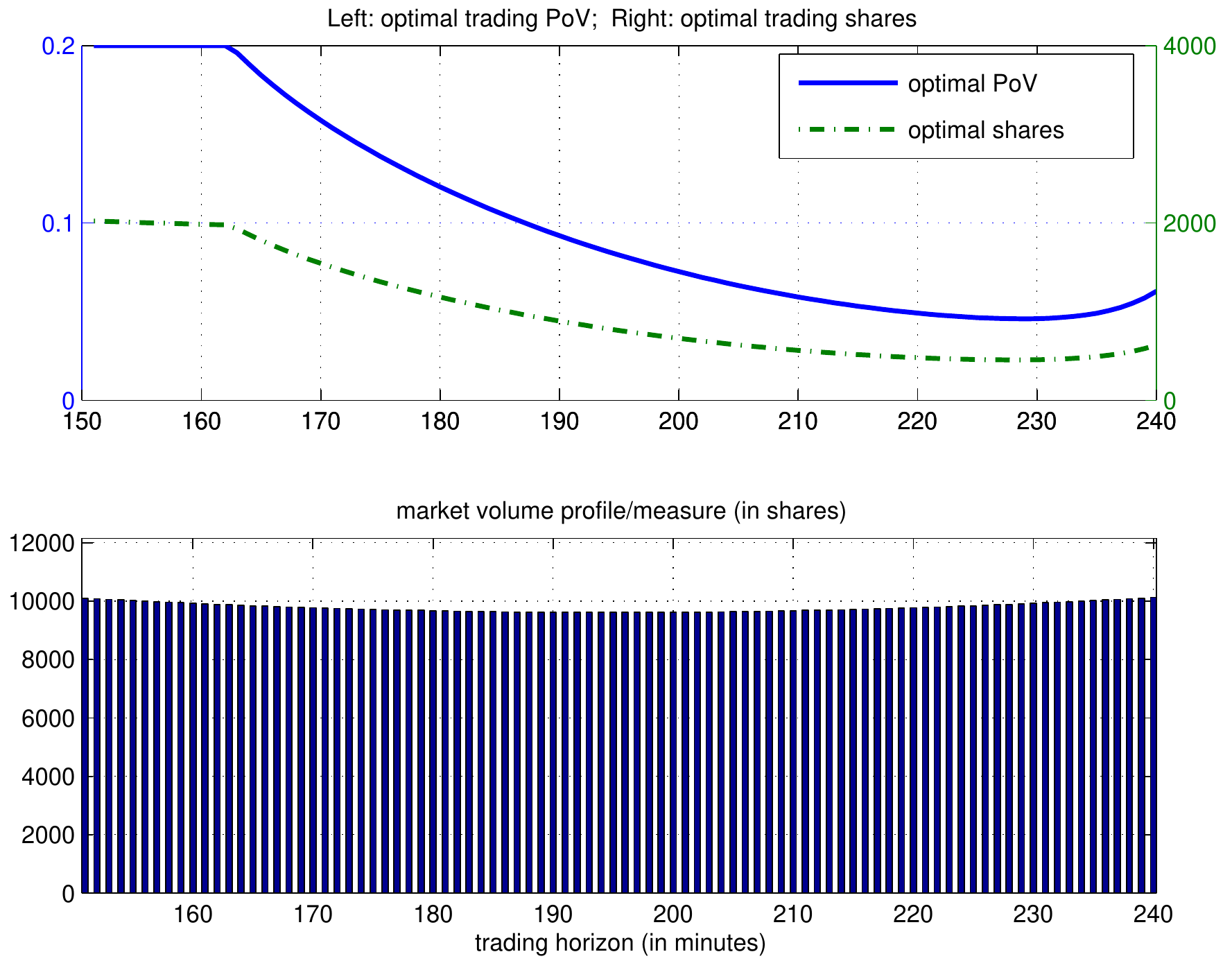}
\caption{ (Effect of Volume Measures) In a U-shaped daily volume measure, volume is relatively low and almost ``flat'' near the hypothetical ``noon'' time (12:00pm - 13:30pm).  Under the given $\maxPoV = 20\%$ and a moderate level of risk aversion, the combined effect of risk aversion and flat volumes encourages to trade faster in the beginning of the horizon. }
\label{F:VS_NearFlat}
\end{figure}

\begin{figure}[h]
\centering
\includegraphics[width=0.51\textwidth]{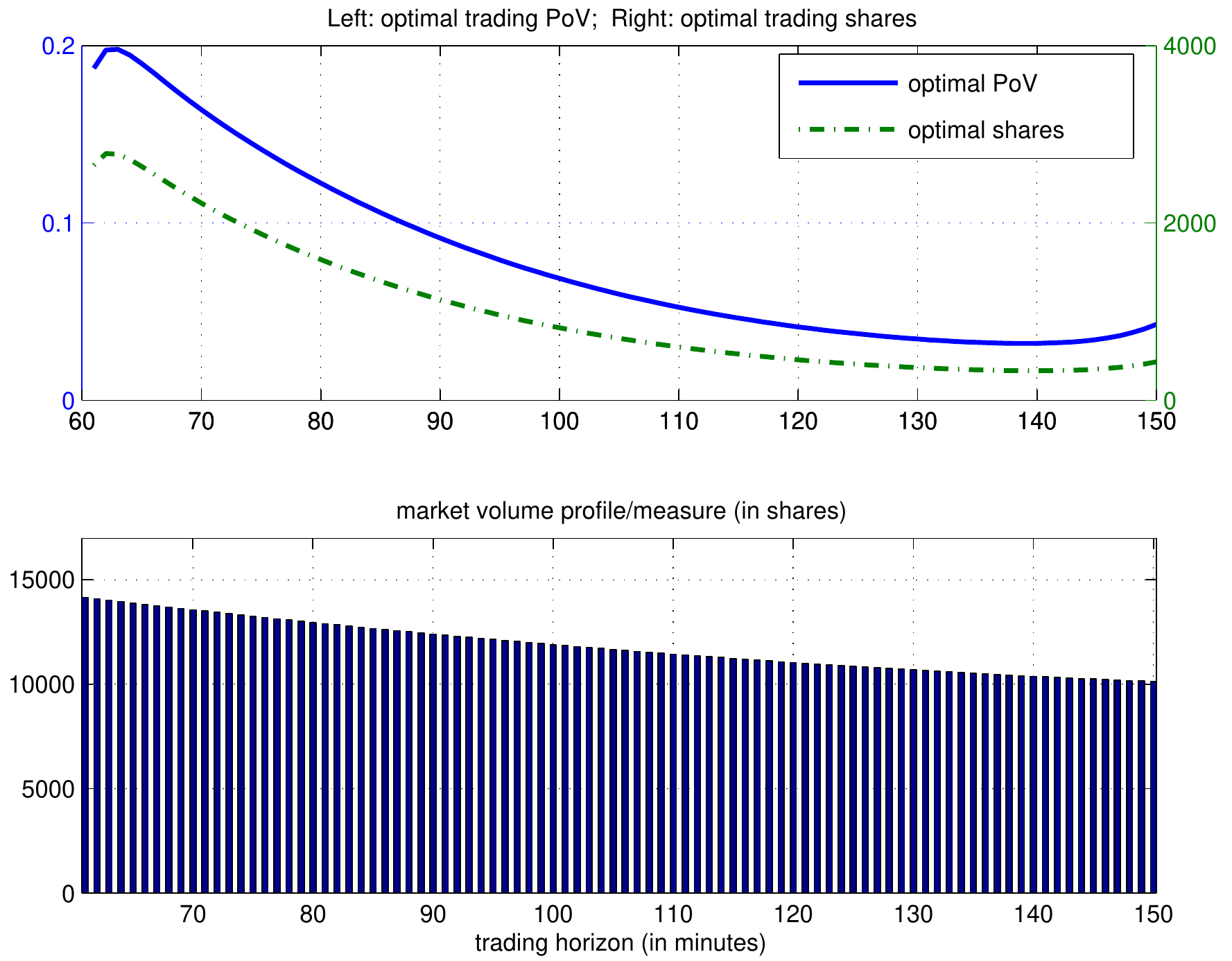} 
\nolinebreak
\includegraphics[width=0.51\textwidth]{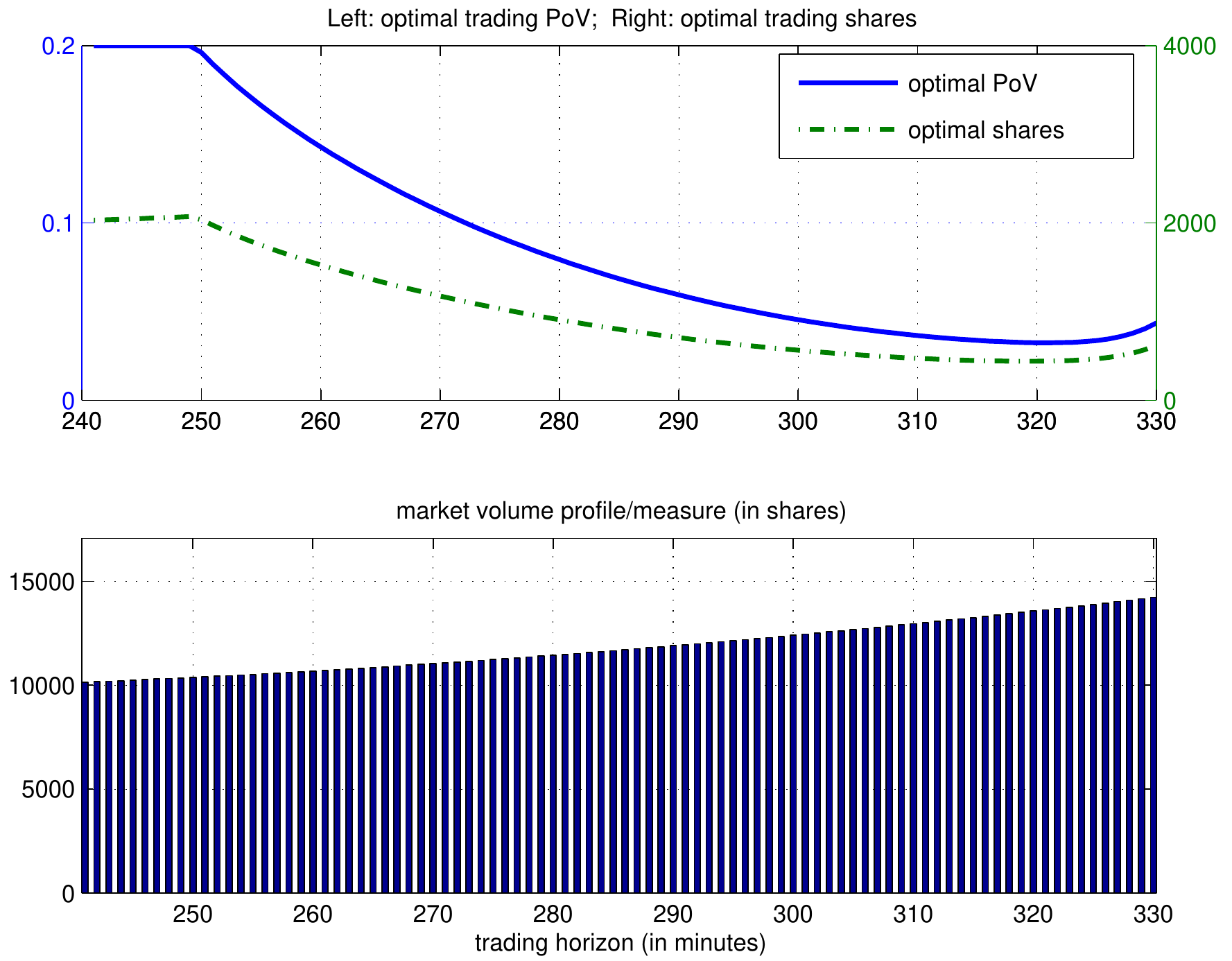}
\caption{(Effect of Volume Measures) Model and trading parameters are being held the same for Fig.~\ref{F:VS_NearFlat} except for the starting time, which affects the volume shape during the trading horizon. Left: trading in the ``morning'' session with more volumes skewed towards the Open; Right: trading in the ``afternoon'' session with volumes more concentrated towards the Close. For the ``afternoon'' session, risk aversion hinders taking full advantage of the richer volumes near the end of the trading horizon, and  encourages instead to participate more in the beginning even the volume is comparatively lower.}
\label{F:VS_UpDown}
\end{figure}

Demonstrated in Fig.~\ref{F:VS_NearFlat} and Fig.~\ref{F:VS_UpDown} are the effects of  volume measures. Under a fixed moderate level of risk aversion ($\lambda=10^{-3}$), trading in a high volume market environment can drive down the average PoV's and hence trading costs (e.g., the left panel of Fig.~\ref{F:VS_UpDown} which simulates typical risk-averse trading in a ``morning'' session). 

\subsubsection{Effect of Cost Components}

\begin{figure}[h]
\centering
\includegraphics[width=0.75\textwidth, height=3in]{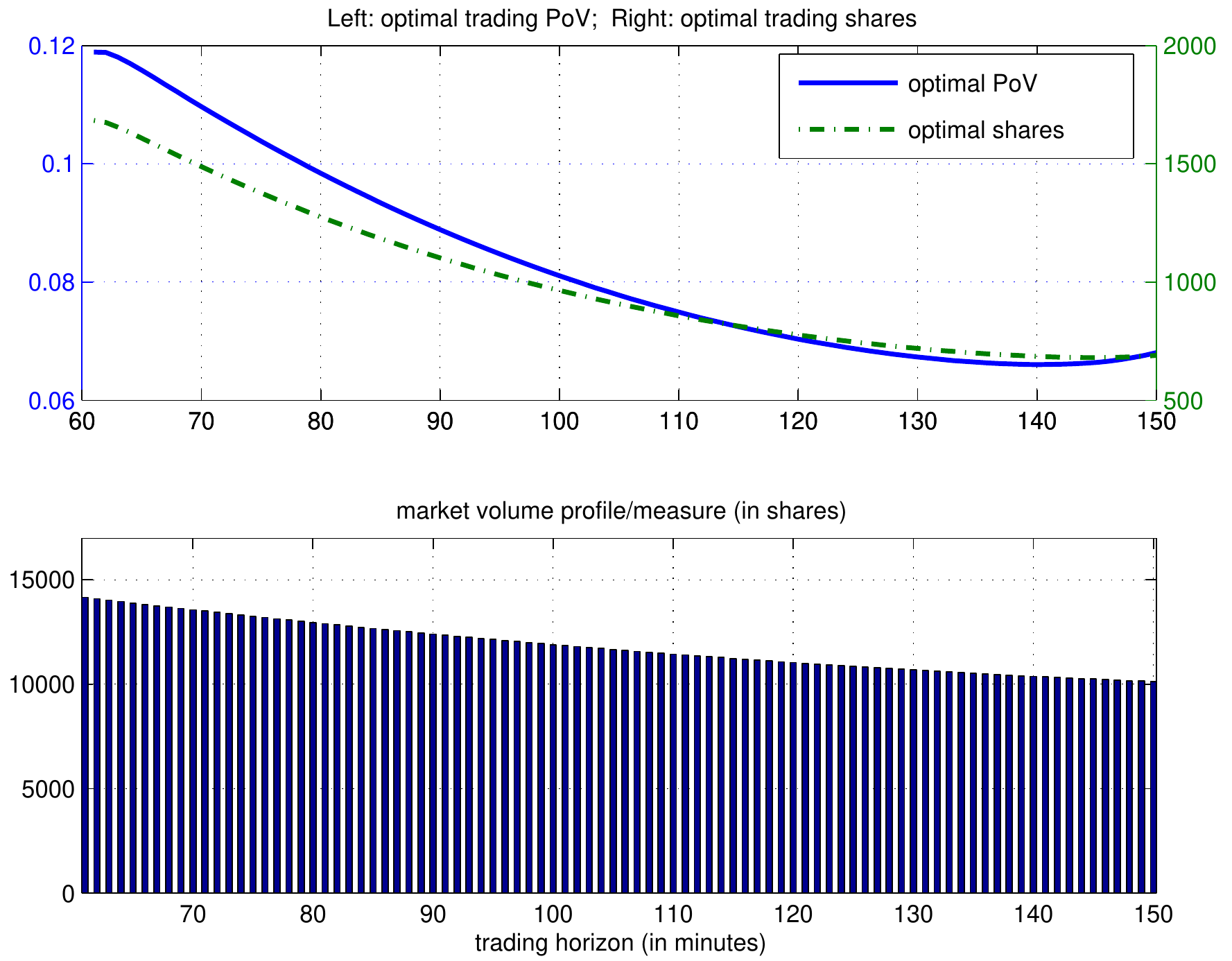}
\caption{ (Effect of Instantaneous Cost) The instantaneous cost  coefficient $\alpha_1^\circ$ is boosted up by 10 times, with $\alpha_2^\circ$ and $\alpha_3^\circ$ fixed at the original moderate level. By definition, the instantaneous cost component is highly localized and trading now vs. later bears no extra penalties. The optimal execution is thus mostly shaped by risk aversion and is typically front-loading. }
\label{F:CC_INST}
\end{figure}

\begin{figure}[h]
\centering
\includegraphics[width=0.75\textwidth, height=3in]{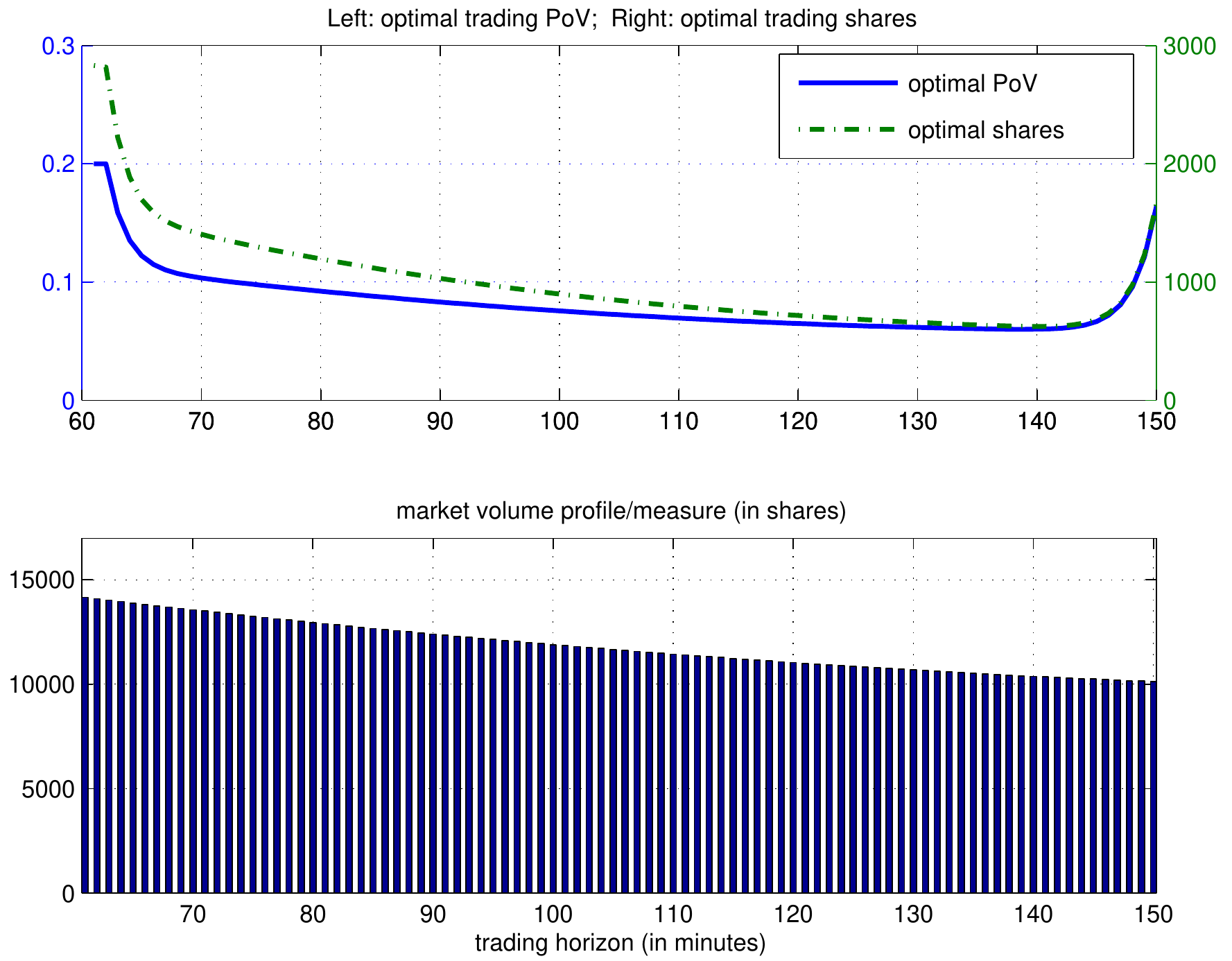}
\caption{ (Effect of Transient Cost)  The transient cost  coefficient $\alpha_2^\circ$ is boosted up by 10 times, with $\alpha_3^\circ$ and $\alpha_1^\circ$ held at the original moderate level. The notable boundary ``wall'' effect arising from the exponential transiency has  been well documented in the classical literature (e.g., see Proposition 2 and 3 in Obizhaeva and Wang~\cite{algo_obizhaeva_wang13}).  }
\label{F:CC_TRAN}
\end{figure}

\begin{figure}[h]
\centering
\includegraphics[width=0.75\textwidth, height=3in]{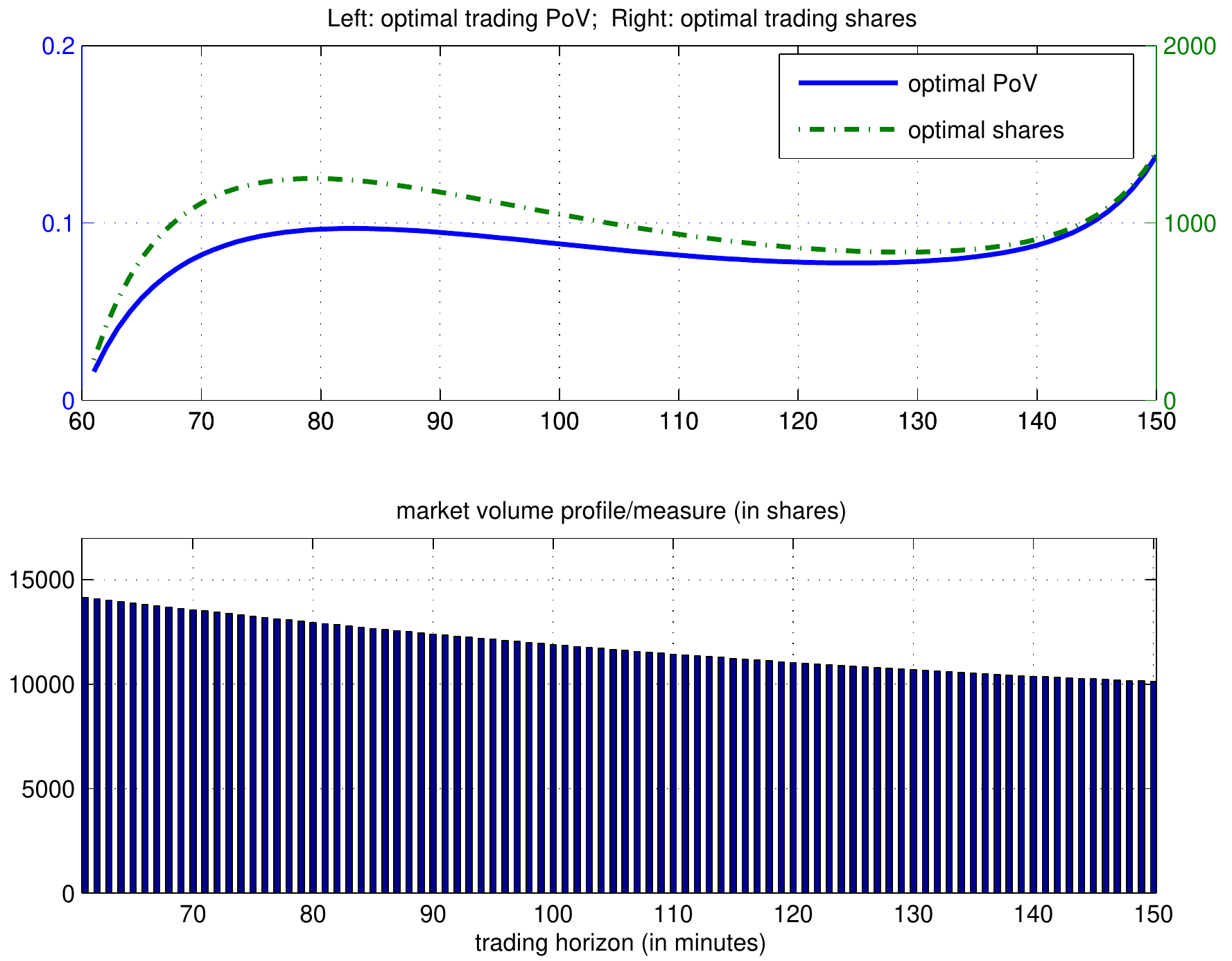}
\caption{ (Effect of Permanent Cost)  The permanent cost  coefficient $\alpha_3^\circ$ is boosted up by 10 times, with $\alpha_1^\circ$ and $\alpha_2^\circ$  held at the original moderate level.  Permanent  cost alone embraces back loading, so that shares traded at later times could pay less permanent cost  built up by the earlier shares. The  optimal trading profile plotted here achieves a balance under the front-loading pressure from risk aversion. }
\label{F:CC_PERM}
\end{figure}

In order to better understand the role of each cost component in the modeling, in the next three figures (Fig.~\ref{F:CC_INST} to~\ref{F:CC_PERM}), we plot optimal solutions corresponding to the boosting of the individual component coefficients: $\alpha_i^\circ, i=1:3$.  For each figure, we have boosted up one of the target coefficients  by 10 times, while holding the other two at the original moderate level. The captions of the individual figures give more details and discussions.

\subsubsection{Effect of Price Dynamics}

In the two panels of Fig.~\ref{F:delta_models}, we also demonstrate the flexibility of the proposed model in dealing with price dynamics other than the classical (linear or geometric) Brownian motions. The left panel plots the optimal solution for the {\em mean-reversal} model, and the right panel for an {\em asymmetric-volatility} model.

\subsubsection*{Mean Reversion}
In the mean reversal model,  one assumes the homogenized price change obeys the following 
equation:
\[
	d \delta_t^\circ = - \kappa \delta_t^\circ dt + \alpha dW_t,
\]
with the initial condition $\delta_{t=0}^\circ = 0.0$. Here $\kappa$ stands for the strength of mean reversal, and $ \alpha dW_t$ for the Brownian component. It can be shown that the solution bears the closed-form:
\[
	\delta_t^\circ = \alpha \cdot \int_0^t e^{-\kappa(t-s)} dW_s,
\]
and that the auto-covariance function is given by
\begin{equation} \label{E:covK_mean_reversion}
	K^\circ_\delta(t, s) 
			=					\frac{\alpha^2}{2\kappa} 
				\cdot 	e^{- \kappa |t -s|}
				\cdot  \left( 1 - e^{-2\kappa(s\wedge t)} \right).
\end{equation}
Notice that for bigger $t, s$,  the covariance function behaves more like an exponential kernel. 

Thus both conceptually and  quantitatively, mean reversion erases long-term memory and only keeps a shift-invariant (when sufficiently away from $t_0=0$) short-term correlation. Unlike Brownian motions for which price uncertainties grow in the order of $\sqrt{t-t_0}$ as time elapses, mean reversion maintains almost a constant level of uncertainty at $\alpha/\sqrt{2\kappa}$.  This implies that faster front-loading trading does not necessarily help reduce risks as risks are time invariant. This is indeed confirmed from the left panel of Fig.~\ref{F:delta_models}. Furthermore, as the kernel asymptotically behaves like the exponential function, we observe the ``wall'' effect at the two boundaries as well documented in the classical literature~\cite{algo_obizhaeva_wang13}.

\subsubsection*{Asymmetric Stochastic Volatility} 

In the next example, we consider a simple form of stochastic volatility:
\begin{equation} \label{E:model:ASV}
	d \delta^\circ_t  = \sigma_0 \cdot \left(  \exp(- \beta \delta^\circ_t \cdot 1_{\delta_t^\circ \leq 0}) \wedge 2.0 \right)  \cdot dW_t,
\end{equation}
where  $\sigma_0$ and $\beta$ are constant. Under this model, instantaneous volatility increases when the price delta is negative, but has been capped under $2\sigma_0$.  In the positive regime when $\delta^\circ_t > 0$, the instantaneous volatility remains at the constant level of $\sigma_0$.  This behavioral  transition has been made possible by the indicator function $1_{\delta_t^\circ \leq 0}$. The volatility is thus asymmetric with respect to the directions of price movements. 

As no closed form exists  for the auto-covariance function, we turn to the Monte-Carlo estimation method using thousands of simulated paths (40,000 in this example). The estimated kernel function is then applied in the proposed execution model. 

\begin{figure}[h]
\centering
\includegraphics[width=0.51\textwidth]{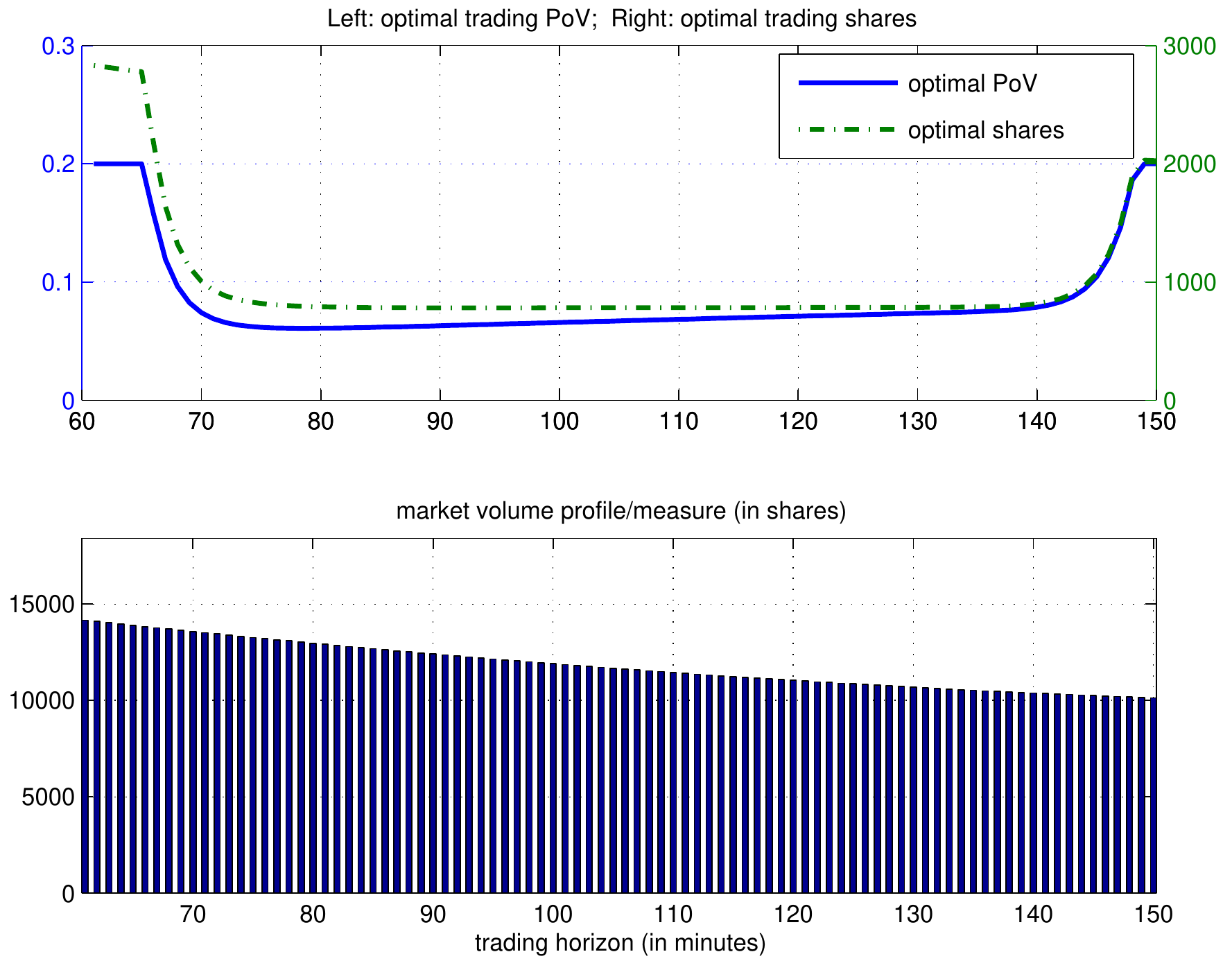} 
\nolinebreak
\includegraphics[width=0.51\textwidth]{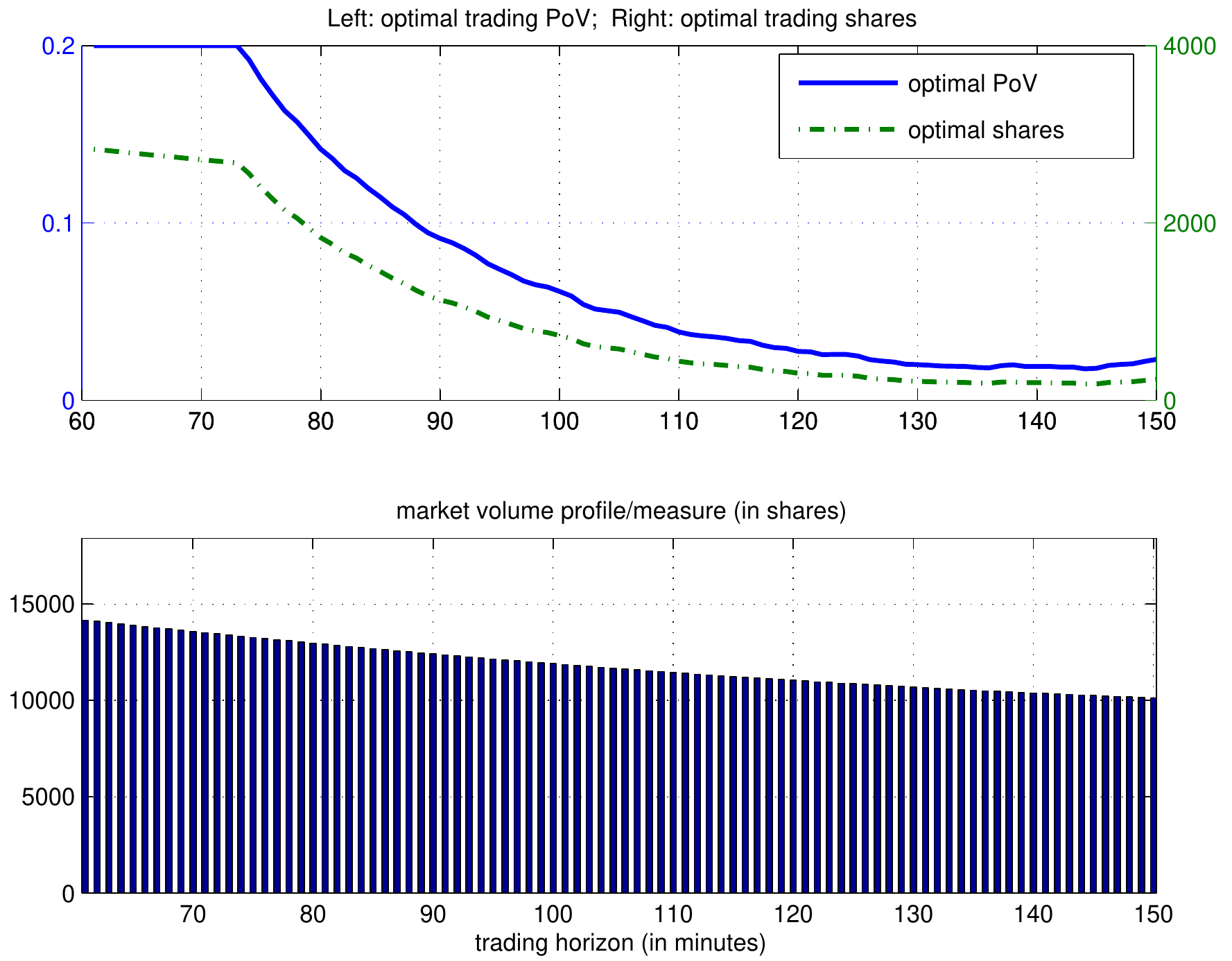}
\caption{(Effect of Price Dynamics) Left Panel: under a mean-reversal (MR) price dynamics; Right Panel: under a price dynamics driven by asymmetric stochastic volatilities (ASV). As the MR dynamics erases long-term memories and converges to time-invariant short-term correlations, execution delay does not necessarily increase risk. Consequently, the front-loading behavior normally associated with risk aversion is weakened. The ``wall'' effect is the classical behavior of the exponential kernel~\cite{algo_obizhaeva_wang13}, which the MR auto-covariance function converges to (see Eqn.~(\ref{E:covK_mean_reversion})).  On the right panel, the asymmetric stochastic volatility model~(\ref{E:model:ASV}) increases the effective instantaneous volatility  compared with the normal  Brownian dynamics and hence the execution risk associated with delaying. Therefore, it enhances the front-loading behavior associated with risk aversion. }
\label{F:delta_models}
\end{figure}

The resulted optimal solution has been plotted in the right panel of Fig.~\ref{F:delta_models}, in which one could observe the non-smoothness associated with the Monte-Carlo kernel  estimation. Since the asymmetry increases the effective volatility, risk aversion enhances the front loading behavior compared with the symmetric case with a constant volatility  $\sigma_0$, which is evident from the plotting.

\section{Conclusion}
While realistic trading models have to be {\em dynamic}, static pre-trade models are still important and widely implemented in the execution industry for a number of reasons and applications  explained in the Introduction section.

The primary objective of the current work has been  to build a continuum model that 
\begin{enumerate}[(a)]
\item  automatically accommodates the broadest varieties of price dynamics,
\item more faithfully engages the role of market volumes in the general reductionism approach of static modeling, 
\item employs impact cost components that bear low complexities but with more market signals embedded and represented, 
\item is versatile enough to allow most popular constraints from clients, and yet
\item is still analytically tractable and computationally feasible.
\end{enumerate}
By working directly with the auto-covariance functions, the model virtually has allowed any price dynamics, in particular, Markovians like Brownians or non-Markovians with memories.
Building upon the foundation of measure theory, we have also treated market volumes as Borel measures over the execution horizons. Pre-trade executions are then considered to be absolutely continuous measures over such measure spaces, which naturally results in the target decision variable to optimize with -- the PoV rate function.  All the four impact cost components have been consistently built upon the PoV function. They are all kept linear but with more market signals integrated in, including volume distances for the transient costs  and cumulative volume normalization  for the permanent costs.   We have also considered heuristically the influence of in-house {\em Child Order Placement} strategies and {\em quantitative market makers} on impact cost building.

In combination, the proposed  pre-trade model has led to a constrained quadratic programming problem in infinite-dimensional Hilbert spaces, which accommodates most linear constraints frequently requested by internal or external clients.  We have in particular worked with the following three primary constraints: (i) monotonicity, (ii) completion, and probably the most important, (iii) PoV capping or volume limits, which is frequently requested from clients. 

We have applied the theories of positive quadratic operators and compact operators in Hilbert spaces to establish both the positivity and compactness of the operators involved, and hence also the existence and uniqueness of optimal executions. One possible numerical scheme for projecting the continuum model onto interval-based grids has also been provided, and several computational examples have been carefully designed to address the effects of all the major factors. 

Despite the versatility of the proposed model, we have to make  some necessary warnings.  Firstly, the model has been primarily designed to be a {\em pre-trade} static model and to become a major service component in the pre-trade packages  offered to both internal or external clients by execution houses. Real {\em dynamic} trading models could be {\em heuristically} built upon such pre-trade models, but have to be integrated with a sound re-optimization  strategy. Next, the current work has been carried out still in the conventional reductionistic approach based on market volumes. With growing analytics and understanding about the {\em limit order book} (LOB) dynamics, it would be naturally interesting to develop LOB based pre-trade models which are both analytically tractable and computationally feasible, and also to compare them with volume-based models (via real trading databases) to quantify the net improvements in pre-trade analytics.

We conclude the work by emphasizing that without the collective academic understanding and industrial practicing started by the many pioneers  and leading practitioners, a portion of whose works have been frequently mentioned, the current work and modeling efforts would be absolutely impossible. 

\section{Disclaimers}

\begin{enumerate}[(1)]
	\item The proposed model has not been the internal or external product of any execution houses where the author had worked. Any potential industrial conflict should be promptly directed to the attention of the author.  
	
	\item Due to the proprietary nature of the industry and the resulting scarcity of real trading data to the public, general expressions like ``based on the experience in the execution houses where the author has worked, ...'', are purely for providing bona fide academic views based on the past working experience with real trading data and results.
	
	\item Any mentioning of certain brand names, e.g., MATLAB, IBM CPLEX Optimizer, or MOSEK, etc., is not a product endorsement from the author for purchasing or investment, but an indication of some popular practices in the contemporary industry. 
	
	\item Execution houses who are interested in the current work and plan to implement it in their systems should be aware of any other operational risks, including for examples, the complexities at market opens or closes, trading halts, stock splitting, or any extreme market events, etc. 

\end{enumerate}

\section{Acknowledgment}
The author wishes to thank the following colleagues on Wall Street: Robert Almgren for explaining to the author, then a freshman on Wall Street in 2007, how Wall Street operates under the introduction of my dear friend Andrea Bertozzi;  Adlar Kim, Max Hardy, Daniel Nehren, Xu Fan, Chang Lin, Jesus Ruiz-Mata, Kathryn Zhao, Calvin Kim, Harry Rana, Arun Rajasekhar, et al., for all the delightful days at J. P. Morgan working  so hard together to build solid Delta One and portfolio hedging, risk, and trading products;  Anlong Li, Iaryna Grynkiv, Paul Radovanovich, Federico de Francisco, Mark Skinner, Merrell Hora, Rishi Dhingra, Lada Kyj,  Nazed Mannan, Allison Greene, Alan Chen, Ajit Kumar, Yihu Fang, Li Xu,  Sunmbal Raza, Peter Ciaccio, Peter Norr, Hasan Ahmed, Bing Song, Ming Yang, Huaguang Feng, and George Liu, et al., for working so closely at the Barclays Capital from day to day as a grand team to deliver high-quality quantitative and technological solutions to modern algorithmic trading and portfolio analytics, and for the favorite donuts and coffees from the Dunkin Donuts, and all mini or max cupcakes from Melissa or Magnolia!

The author is particularly grateful to these colleagues and close friends, who had not only shaped his wall street career but also deeply and warmly touched his everyday life: Adlar Kim, Max Hardy, Daniel Nehren, Paul Radovanovich, Bing Song, and Huaguang Feng.

%\bibliographystyle{plain}
%\bibliography{bibAlgoTrading_JShen}

\end{document}